# Novel Magnetic Materials for Spintronic Device Technology

Aakanksha Sud, Akash Kumar and Murat Cubukcu

**Abstract**

Spintronics, a transformative field of research, leverages the spin of electron to revolutionize electronic devices, offering significant advantages over traditional charge-based systems. This chapter highlights the critical role of novel magnetic materials in advancing spintronic technologies by addressing their fundamental properties, fabrication methods, and applications. A diverse range of materials, including ferromagnetic, antiferromagnetic, non-collinear antiferromagnetic, synthetic antiferromagnetic, ferrimagnetic, and multiferroic systems, is explored for their unique contributions to spintronic devices. Advanced fabrication techniques, such as bulk crystal formation, thin-film deposition, and nano-structuring, are detailed along characterization methods including spin-orbit torque analysis and magnetization dynamics studies. Emerging research on three-dimensional spin textures, domain walls, skyrmions, and magnons highlighted for its potential for novel spintronic applications. Additionally, the chapter reviews applications in memory technologies, spintronic nano-oscillators, and neuromorphic computing. Finally, it concludes by examining future research directions, challenges, and opportunities in spintronics, providing insights into breakthroughs that are shaping the future of this rapidly evolving field.

**Keywords**: Spintronics, magnetic materials, spin-orbit torque, skyrmions, magnonics, memory technologies, neuromorphic computing

*Aakanksha Sud\**
*Frontier Research Institute for Interdisciplinary Sciences (FRIS), Tohoku University, 6-3 Aramaki, Aoba-ku, Sendai, 980-0845, Japan and also with Research Institute of Electrical Communication (RIEC), Tohoku University, 2-1-1 Katahira, Aoba-ku, Sendai, 980-8577 Japan.*
*(\*e-mail: sud.aakanksha.c3@tohoku.ac.jp)*
*Akash Kumar*
*Department of Physics, University of Gothenburg, 41296, Gothenburg, Sweden and also with*
*Center for Science and Innovation in Spintronics (CSIS), Tohoku University, 2-1-1 Katahira, Aoba-ku, Sendai, 980-8577 Japan.*
*Murat Cubukcu*
*National Physical Laboratory, Teddington, TW11 0LW, United Kingdom and also with University College London, Gower Street, London WC1E 6BT, United Kingdom*

# Table of Contents

## 1.1 Introduction: Overview of spintronics

Spintronics, or spin-based electronics [1], has emerged as a transformative field in modern technology. Originating in the 20th century [2], it has rapidly evolved, driving a paradigm shift in information processing and memory technologies. Unlike traditional electronics, which relies solely on the electron's charge, spintronics exploits the electron's intrinsic angular momentum—its spin—as an additional functional parameter. This enables entirely new approaches to device design, performance enhancement and optimization [3, 4]. The growing interest in spintronics stems from its ability to deliver substantial advantages, including reduced power consumption, faster operation speeds, and increased device density. These features make spintronic systems particularly appealing for next-generation computing and data storage solutions [5].

Quantum computing [6] has further amplified the importance of spintronics by leveraging spin polarization, which retains coherence longer than charge polarization, making it ideal for qubits. Pioneering work by Awschalom *et al*. (1999) [7] and Kikkawa *et al*. [8] demonstrated spin coherence manipulation and long spin lifetimes in semiconductors, laying the foundation for spin-based quantum computing.

Today, spintronic devices such as spin-transfer torque magnetic random-access memory (STT-MRAM) [9] and magnetic tunnel junctions (MTJs) [10] have evolved from experimental concepts to commercial product [11]. These innovations have reshaped the technology landscape, offering non-volatile, high-performance solutions for data processing and storage [12]. Furthermore, the advancements in spintronics provide a solid foundation for future breakthroughs in quantum computing, neuromorphic systems [13,14,15,16], and beyond. These developments are fundamentally driven by the unique properties of magnetic materials, which are essential for enabling the precise control and manipulation of spin dynamics in these devices.

**Importance of Magnetic Materials**

At the core of spintronics lies the interplay of magnetic materials with electron spin which serve as the foundation for spintronic devices [17]. Their unique magnetic properties, such as anisotropy, coercivity, and saturation magnetization—determine the efficiency and functionality of spintronic systems. Materials like ferromagnets, antiferromagnets, and ferrimagnets exhibit distinct spin dynamics and coupling mechanisms, enabling a wide range of spintronic applications [18]. Recent advances in novel materials, including two-dimensional magnets [19], non-collinear antiferromagnets [20], and topological insulators [21], have opened new avenues for innovation in the field. These materials not only improve device performance but also drive the development of emerging applications, such as neuromorphic computing and quantum information processing [22].

This chapter explores the critical role of magnetic materials in advancing spintronic technologies. It begins by highlighting the fundamental properties and classifications of magnetic materials, underscoring their significance in spintronics. The discussion then extends to various types of magnetic materials, including ferromagnetic, antiferromagnetic, non-collinear antiferromagnetic, ferrimagnetic, and multiferroic systems. These materials are examined with respect to their unique characteristics, applications, and relevance to spintronic devices. The chapter also explores cutting-edge fabrication and characterization techniques that enable the precise engineering of magnetic materials for specific applications. Emerging areas of research, such as three-dimensional spin textures [23,24], magnonic systems [25,26], and spin-orbit torque materials [27], are reviewed, offering insights into the future of spintronic devices. By the end, readers will gain a comprehensive understanding of how magnetic materials form the foundation of spintronics, driving innovations in memory technologies, unconventional computing [28], and next-generation quantum systems [29,30].

**1.2 Types of magnetic materials**

Magnetic materials are essential to spintronic devices, providing the means to control and manipulate electron spin. Each category, from ferromagnets with strong spontaneous magnetization [31] to antiferromagnets with antiparallel spin alignments, offers unique properties tailored to specific applications. Emerging materials, such as non-collinear antiferromagnets and two-dimensional ferromagnets, have further expanded the possibilities of spin-based technologies. This section explores the principles, examples, and applications of these key magnetic materials, emphasizing their critical roles in advancing spintronics.

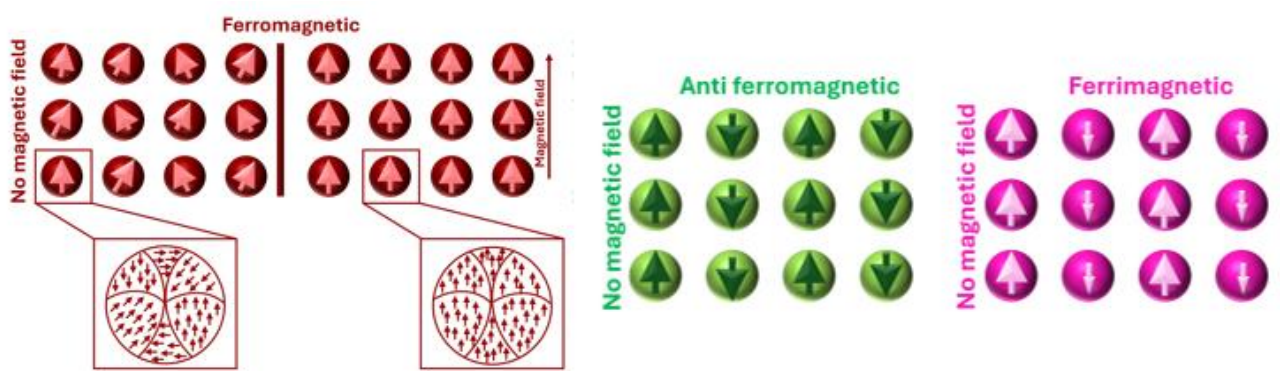

Figure 1.1: Illustration of magnetic dipole arrangement in different magnetic material.

**Ferromagnetic materials**

Ferromagnetism, one of the most well-known and studied magnetic phenomena, was first observed in natural materials like lodestone (magnetite, $Fe_3O_4$) with its magnetic properties described as early as 600 BCE by ancient Greeks [32] and systematically studied in 1600 by William Gilbert in his work *De Magnete* [33]. Later, ferromagnetism was scientifically examined in metals such as iron (Fe), cobalt (Co), and nickel (Ni) and their alloys. These materials exhibit spontaneous magnetization, where magnetic moments within microscopic regions, called domains, align parallel to one another as shown in Fig. 1.1 (a), even in the absence of an external magnetic field. This behaviour arises from quantum mechanical exchange interactions [34] which energetically favour parallel spin alignment among neighbouring electrons, particularly in the partially filled *d*-orbitals of these transition metals. Properties such as magnetic anisotropy, coercivity, and the Curie temperature influence their behaviour and define their utility in technological applications [35].

Ferromagnetic materials, particularly Fe, Co and Ni, have been indispensable in traditional applications such as magnetic cores, sensors, and recording devices [36]. Fe, with its high saturation magnetization, is widely used in transformers and electrical generators. Co known for its high Curie temperature and strong anisotropy, plays a crucial role in aerospace and high-performance magnetic alloys [37]. Ni, on the other hand, is commonly used in thin films and multilayers for spintronics owing to its ductility and moderate magnetization [38]. These materials have been studied extensively in foundational works such as Bozorth's "*Ferromagnetism*" (1947) [39], which provides a comprehensive experimental analysis, and Morrish's "*The Physical Principles of Magnetism*" (1967) [40], which explores the theoretical aspects of their magnetic properties and behavior. For a detailed exploration of these materials, readers may refer to these seminal works. Beyond traditional materials, rare-earth elements such as gadolinium (Gd) and lanthanum (La) [41] exhibit unique magnetic properties that make them valuable for spintronic applications. Gd [42], with its Curie temperature of approximately 294 K, is particularly notable for its near-room-temperature ferromagnetic behavior and high spin polarization, making it a promising candidate for spin-filter devices and magneto-optical applications [43, 44].

However, terbium (Tb) surpasses Gd in magnetization at low temperatures due to its strong magnetocrystalline anisotropy [45] and larger 4*f* orbital contribution [41], which enhances its potential for high-coercivity spintronic devices. Dysprosium (Dy), with its even stronger spin-orbit coupling [41], is also being explored for advanced magneto-optical and spintronic applications that require robust control over magnetic anisotropy [46].

Recent advancements have significantly expanded the scope of ferromagnetic materials, introducing novel systems with enhanced properties. Two-dimensional (2D) ferromagnets such as $Cr_2Ge_2Te_6$, $CrI_3$, and CrSBr [47,48] exhibit stable magnetic ordering at the monolayer scale, driven by strong spin-orbit coupling and anisotropic interactions, making them ideal for miniaturized spintronic devices [49]. The emergence of van der Waals (vdW) ferromagnets has further broadened the possibilities in spin-orbit torque (SOT) applications, offering high-quality interfaces and tunable magnetic anisotropy. Ostwal *et al*. [50] demonstrated SOT-induced magnetization switching in $Cr_2Ge_2Te_6$, while Wang *et al*. [51] and Alghamdi *et al*. [52] achieved similar switching in $Fe_3GeTe_2$, highlighting the potential of vdW ferromagnets for ultra-low-power spintronic devices. Additionally, MacNeill *et al*. [53] and Shi *et al*. [54] explored magnetization switching in $WTe_2$/ferromagnet heterostructures, leveraging the Edelstein effect to convert charge currents into spin currents and achieved a critical switching current density of $2.96 \times 10^5$ A cm$^{-2}$ [54]. Topological insulators (TIs) play a crucial role in enhancing charge-to-spin conversion efficiency due to spin-momentum locking [55,56]. Mellnik *et al*. [57] first observed spin current generation via the Rashba-Edelstein effect in $Bi_2Se_3$, while Fan *et al*. [58] demonstrated magnetization switching in $(Bi_{0.5}Sb_{0.5})_2Te_3$ heterostructures, reporting exceptionally high spin Hall angles. Khang *et al*. [59] further optimized this efficiency by achieving ultralow-power SOT switching in a BiSb/MnGa bilayer, reaching a critical current density of $1.5 \times 10^6$ A cm$^{-2}$. Recognizing scalability challenges in molecular beam epitaxy-grown TIs, Mahendra *et al*. [60] developed an industry-friendly sputtered BiSe/CoFeB bilayer, reducing the switching current to $4.3 \times 10^5$ A cm$^{-2}$, making TIs more viable for commercial spintronic applications.

The combination of 2D ferromagnets, topological insulators, and transition metal dichalcogenides (TMDs) [61] is shaping the future of SOT-driven spintronic devices. Their unique properties enable efficient spin-charge interconversion, miniaturization, and low-power operation, paving the way for next-generation spintronic memory and logic architectures with enhanced scalability and energy efficiency. Magnetic Weyl semimetals (WSMs) [62] exhibit unique topological features, including Weyl points, Fermi arcs, and chiral anomaly-induced negative magnetoresistance [63,64], making them highly promising for spintronic applications. Several Heusler compounds have been identified as potential Weyl semimetal (WSM) candidates, including Co-based XCo$_2$Z (X = Nb, Zr, V, Ti, Hf; Z = Sn, Ge, Si) [65], as well as strained Heusler topological insulators [66]. The antiferromagnetic (AFM) Heusler compound CuMnAs [67] (*see next section for details about antiferromagnetic materials*) exhibits non-symmorphic symmetry-protected Dirac semimetal (DSM) behavior, while chiral AFM Heusler materials Mn$_3$X (X = Sn, Ge) [68] display a strong anomalous Hall effect (AHE) at room temperature and have been predicted as AFM WSMs. Additionally, Heusler-like ternary compounds such as ZrSiS [69] and LaAlGe [70] further expand the family of topological semimetals, emphasizing the growing potential of Heusler materials in spintronics. In 2019, Belopolski *et al*. [71] provided direct experimental evidence for the topological nature of Co$_2$MnGa, identifying bulk Weyl Fermi arcs through ARPES measurements. Their study confirmed Co$_2$MnGa as a magnetic Weyl semimetal, characterized by topologically protected surface states. Co$_2$MnGa exhibits a large anomalous Hall effect (AHE) [72], arising from its topologically protected surface states and Berry curvature [72]. Heusler alloys, such as Co$_2$FeSi and NiMnSb [73], are critical materials in spintronics due to their high spin polarization and tunable electronic properties, making them essential for applications like spin valves and MTJs. Their half-metallic nature facilitates efficient spin transport, which is crucial for enhancing the performance of MRAM. Additionally, defect-induced ferromagnetism in oxides like ZnO, $HfO_2$, and $TiO_2$ [74] opens up alternative pathways for integrating magnetism into unconventional platforms, thereby expanding the material landscape for spintronic applications.

The applications of ferromagnetic materials extend beyond spintronics into diverse fields. They form the foundation of devices like giant magnetoresistance (GMR) structures, STT-MRAM [75], and advanced magnetic sensors. In biomedicine, ferromagnetic nanoparticles are used for targeted drug delivery [76], magnetic hyperthermia in cancer treatment, and as contrast agents in magnetic resonance imaging (MRI) [77]. These materials also play a critical role in neuromorphic computing [78], where ferromagnetic memristors and synapses enable energy-efficient artificial neural networks. Furthermore, their integration into quantum technologies, such as hybrid quantum systems, highlights their versatility in next-generation computing and information storage [79].

**Antiferromagnetic Materials**

Antiferromagnetic (AFM) materials, characterized by the compensation of magnetic moments across opposing sublattices as shown in Fig. 1.1(b), have evolved from theoretical curiosity to essential components in spintronics. Once described as "*interesting but useless*" by Louis Néel in his Nobel lecture [80], these materials have since demonstrated unique advantages, including the absence of stray fields, terahertz-scale magnetization dynamics, reduced crosstalk and robust thermal stability under external magnetic fields [81]. These properties make AFMs invaluable for advanced spintronic applications, particularly in densely packed device environments where minimizing magnetic interference is crucial [82].

The first major application of AFMs in spintronics emerged in the early 1990s with the development of spin valve structures based on GMR [83], a breakthrough discovered by Baibich *et al*. [84] and Binasch *et al*. in 1988 [85]. This innovation revolutionized hard disk drive technology by enabling highly sensitive read heads, where an AFM layer was used to pin the magnetization of one ferromagnetic (FM) layer via exchange bias [86]. This configuration significantly enhanced data storage density and stability, highlighting the crucial role of AFMs in practical spintronic devices. Today, AFMs are integral to MRAM technologies [87], including toggle-MRAM [88], STT-MRAM [89], and spin-orbit torque (SOT)-MRAM [90], where they stabilize MTJs [91] and ensure reliable data storage and retrieval.

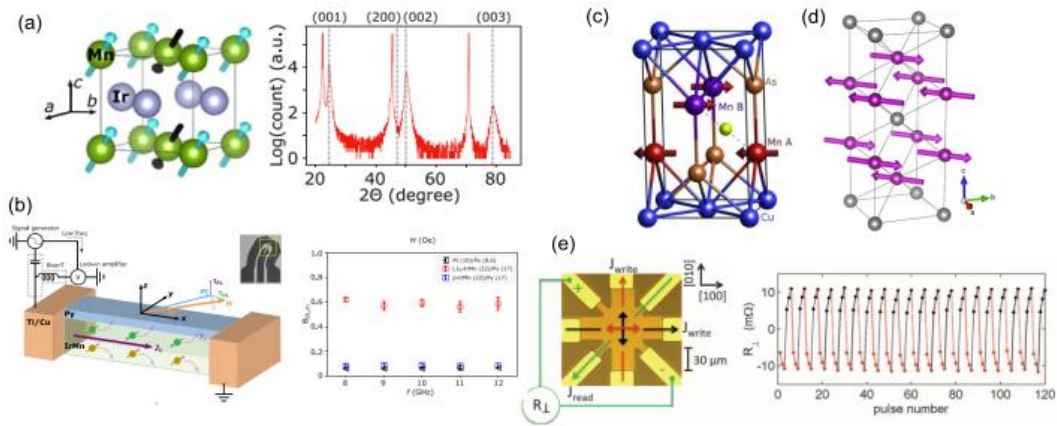

Fig 2. Schematic drawing of L10-IrMn unit cell along with the XRD $\theta$-$2\theta$ scan of IrMn along the (001) direction. The dotted lines correspond to the reference peak positions of bulk L10-IrMn. (b) Measurement setup using an IrMn bilayer to obtain SOT efficiency. Top right shows the optical image of the device and electrode (dark color). The Right panel shows the SOT efficiency of L10-IrMn, p-IrMn, and Pt. Figure adapted from Zhou *et al*. [93]. (c, d) Crystal structure for CuMnAS (c) and $Mn_2Au$ (d) with AFM ordering. The two Mn spin-sublattices A and B (red and purple) are inversion partners in CuMnAS and The Mn moments form ferromagnetic sheets perpendicular to c direction in Mn2Au. Figure adapted from Wadley *et al* [97] and Barthem *et al* [101]. (e) Electrical switching in CuMnAs. Optical microscopy image of the device and schematic of the measurement setup. (c) Transverse resistance changes after three successive writing pulses along [100] and [010] axes of CuMnAs. The reading current is applied along [110], with signals recorded 10 s post-pulse after offset subtraction [97].

Collinear antiferromagnetic materials [92], where magnetic moments align in strictly antiparallel configurations, have been at the forefront of spintronic research. Materials such as IrMn [93], CuMnAs [94], and $Mn_2Au$ [95] exhibit remarkable properties, including anisotropic magnetoresistance (AMR) and spin Hall magnetoresistance (SMR). These effects enable the electrical manipulation and detection of the Néel vector, providing robust methods for data storage in MRAM devices. IrMn, was identified for its exchange bias effect in spin valves by Devasahayam *et al*. in 1999 [96]. Figure 2 (a) shows the crystallographic structure for IrMn.

In 2019 Zhou *et al.* [93] demonstrated that L1$_0$-ordered IrMn, a collinear antiferromagnet, exhibits enhanced SOT efficiency up to $0.60 \pm 0.04$, as shown in Figure 2 (b), making it a promising candidate for energy-efficient spintronic applications. Recent breakthroughs have enabled electrical switching of the Néel vector in CuMnAs (Wadley *et al*., 2016) [97] and Mn$_2$Au (Meinert *et al*., 2018) [98] through SOTs, paving the way for ultrafast, energy-efficient devices. Figure 2 (c,d) depicts crystal structure in CuMnAs and Mn$_2$Au both of which possess a similar crystal structure, with Mn sites forming locally non-centrosymmetric inversion partners.

The electrical switching in CuMnAs (Figure 2 (e)) was demonstrated by Wadley *et al*. [97], who achieved reversible 90° Néel order switching via current injection, with Mn moments aligning perpendicular to the current, detectable through AMR.These innovations enable manipulation of antiferromagnetic domains at terahertz frequencies as demonstrated in Mn$_2$Au by Bodnar *et al*. in 2018 [99]. This is significantly faster than ferromagnetic counterparts, while their thermal stability ensures reliable operation under extreme conditions. Additionally, materials such as CoFeB/IrMn [100] leverage exchange bias interactions to achieve deterministic field-free magnetization reversal, expanding the scalability and energy efficiency of AFM-based memory technologies.

In contrast to collinear systems, noncollinear antiferromagnetic materials [102] exhibit complex triangular or kagome spin arrangements as in Mn$_3$X (X = Sn, Ge, Ga), which break conventional symmetries and give rise to exotic transport phenomena. Figure 3 (a-c) shows the crystal structure of $Mn_3X$ [103]. In these materials, Mn moments in neighbouring Kagome bilayers form cluster magnetic octupoles, creating a ferroic order, as depicted in Figures 3(b) and 3(c), within the inverse triangular spin structure. The presence of long-range triangular antiferromagnetic order breaks macroscopic time-reversal symmetry, a characteristic typically associated with ferromagnets. When combined with spin-orbit coupling (SOC), this leads to a nonzero k-space Berry curvature, resulting in a large AHE similar to what is observed in ferromagnets.

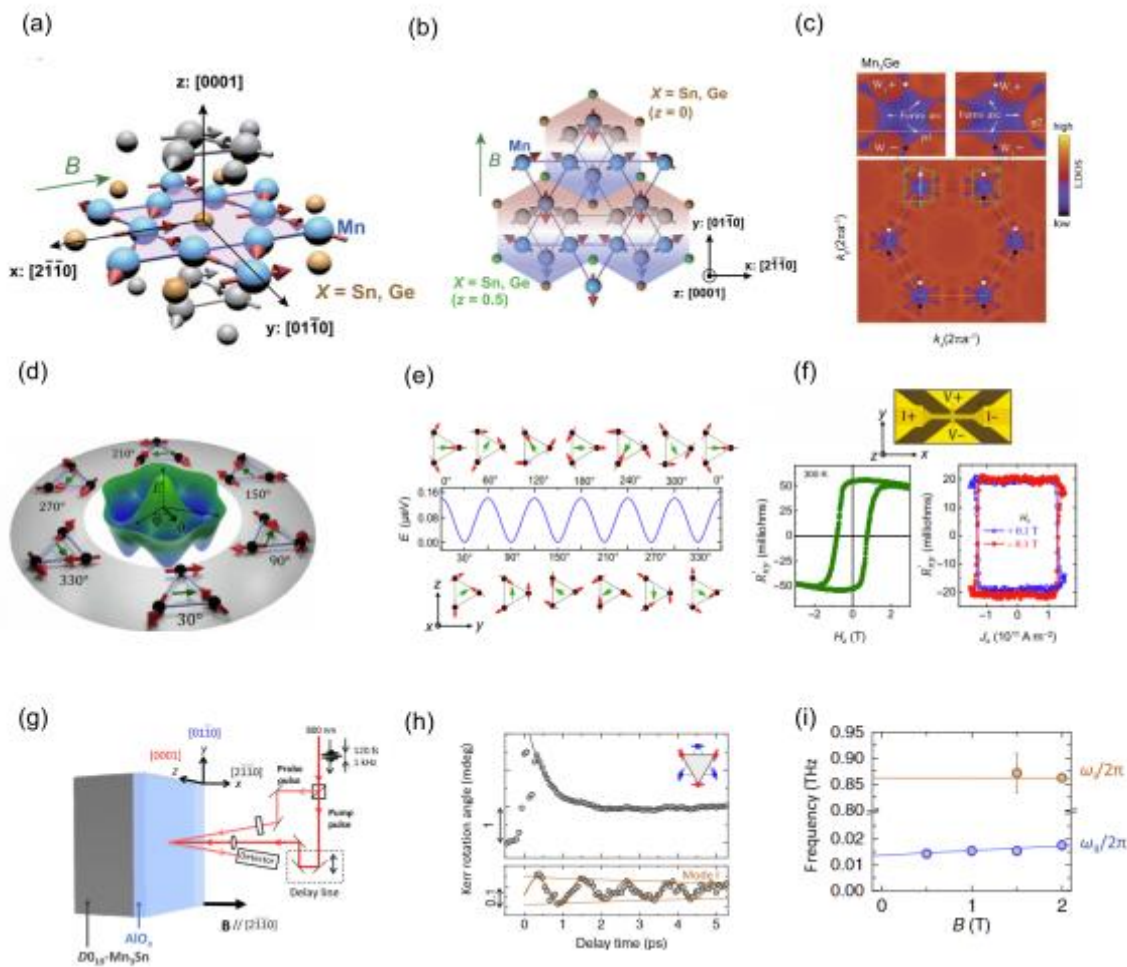

Fig 3. (a) Crystal structure and electronic of $Mn_3X$ (X = Sn, Ge). Mn atoms form a kagome lattice in the hexagonal ab-plane, exhibiting an inverse triangular spin order (purple). (b) Under an applied magnetic field B || y, Mn moments align into cluster magnetic octupoles (colored hexagons), representing a ferroic spin structure. (c) Surface local density of states (LDOS) at the Fermi level in $Mn_3Ge$, showing Weyl points (white and black dots) and associated Fermi arcs ($p_1$ and $p_2$). (a-c) adapted from Chen *et al*. [103]. (d) Energy landscape of the sixfold degenerate magnetic states in $Mn_3Sn$ with net moment rotation. (e) Degenerate ground states and 2D energy landscape for $\varphi$ variation. (f) Anomalous Hall resistance (left) of $Mn_3Sn$/W/TaN at 300 K and current-induced switching (right) under $H_x = \pm 0.1$ T, with a Hall bar micrograph showing field and current directions. (d-f) adapted from [114]. (g) Set-up for Time-resolved MOKE measurements of $Mn_3Sn$ under a 2T field (h) Cluster magnetic octupole oscillations obtained from Background-subtracted TR-MOKE signals reveal resonant

frequencies of 0.86 THz and 18 GHz with damping constants α = 0.02 and 1.0, respectively. (i) Magnetic field dependence of oscillation modes. (g-i) are reprinted from [115].

The AHE in noncollinear antiferromagnets was theoretically predicted in 2014 [104,105] and experimentally confirmed in Mn$_3$Sn in 2015 by Nakatsuji *et. al* [106]. The AHE in Mn$_3$Sn exhibited anomalous Hall conductivity reaching ~20 Ω$^{−1}$ cm$^{−1}$ at room temperature, which is comparable to some ferromagnetic metals, despite these materials having a negligible net magnetization due to spin canting. This discovery confirmed the topological nature of the AHE [107], rooted in Berry curvature and challenged the conventional belief that AHE scales with magnetization. Similarly, Mn$_3$Ge [108] also exhibited a significant AHE, owing to its noncollinear spin texture. These noncollinear systems offer unique functionalities, such as the magnetic spin Hall effect (MSHE), where spin currents are generated orthogonally to both the current flow and the Néel vector via spin-orbit coupling. In 2017, Železný et al. [109] predicted the MSHE in noncollinear antiferromagnets, demonstrating that spin currents can be generated without the need for SOC. This effect was experimentally verified in Mn$_3$Sn [110] and Mn$_3$Ir [111], enabling field-free magnetization switching in adjacent ferromagnetic layers, significantly enhancing device efficiency. Additionally, the anomalous Nernst effect [112,113] has been observed, further expanding their potential for thermoelectric applications.

In 2022 Pal *et al*. [114] reported a previously unobserved seeded spin-orbit torque (SSOT) effect, in which current can manipulate the magnetic states of even thick layers of the Mn$_3$Sn. They demonstrated current induced switching in Mn$_3$Sn which cannot be accounted for by conventional SOT. Mn$_3$Sn features six energetically degenerate magnetic ground states, each characterized by the net moment ***m*** aligning along specific polar angles φ (Figure 3(d)). Unstable equilibrium states occur at φ = 60°, 120°, …, 360°, corresponding to local energy maxima (Figure 3(e)). These magnetic configurations, along with their energy barriers, govern the SSOT switching mechanism (Figure 3(f)).

Time-resolved optical studies have revealed cluster octupole oscillations in $Mn_3Sn$ [115] (Figure 3 (g-i)), supporting ultrafast spin dynamics with switching timescales below 10 ps, making these materials highly suitable for high-speed spintronic applications. The SOT efficiency in these materials exceeds that of conventional heavy metal systems like Pt and Ta, positioning them as attractive candidates for memory and logic applications.

In 2020, Tsai *et al*. [116] demonstrated that the triangular spin order in $Mn_3Sn$ could be manipulated using SOTs with a small in-plane field of ~0.1 T. This achieved a critical switching current density comparable to ferromagnetic SOT devices. Furhermore, the AHE in $Mn_3Sn$ and $Mn_3Ge$, which is tied to the topological properties of Weyl nodes, allows for electrical detection and control of magnetic order, making these materials promising candidates for spintronic applications. Recent studies have also explored strain engineering in multiferroic heterostructures [117], offering an energy-efficient method to manipulate noncollinear antiferromagnetic order. The unique combination of ultrafast spin dynamics, high stability, and tunable spin textures positions these materials at the forefront of emerging spintronic and neuromorphic computing technologies, including racetrack memory and topological quantum devices [102].

**Synthetic antiferromagnetic (SyAF) materials**, consisting of alternating ferromagnetic layers separated by nonmagnetic spacers, provide a versatile and tunable alternative to natural antiferromagnets [118] as illustrated in Figure 4(a). Unlike their crystalline counterparts, where exchange interactions are strictly defined by atomic arrangements, SyAFs utilize the Ruderman–Kittel–Kasuya–Yosida (RKKY) interaction [118]. This oscillatory exchange coupling allows for precise control of magnetic configurations by varying the spacer thickness or material composition. The ability to tailor the magnetic properties of SyAFs makes them highly valuable for spintronic applications. A key distinction between SyAFs and natural antiferromagnets lies in their exchange interactions and antiferromagnetic order. In SyAF, the exchange coupling is weaker and tunable, allowing for easier manipulation of magnetic states. While natural antiferromagnets exhibit atomic-scale alternation of magnetic moments, SyAFs maintain long-range antiferromagnetic order across nanometer-scale ferromagnetic layers. This extended order facilitates the use of semiclassical models for describing electron transport, making SyAFs well-suited for conventional magnetic sensing and memory technologies.

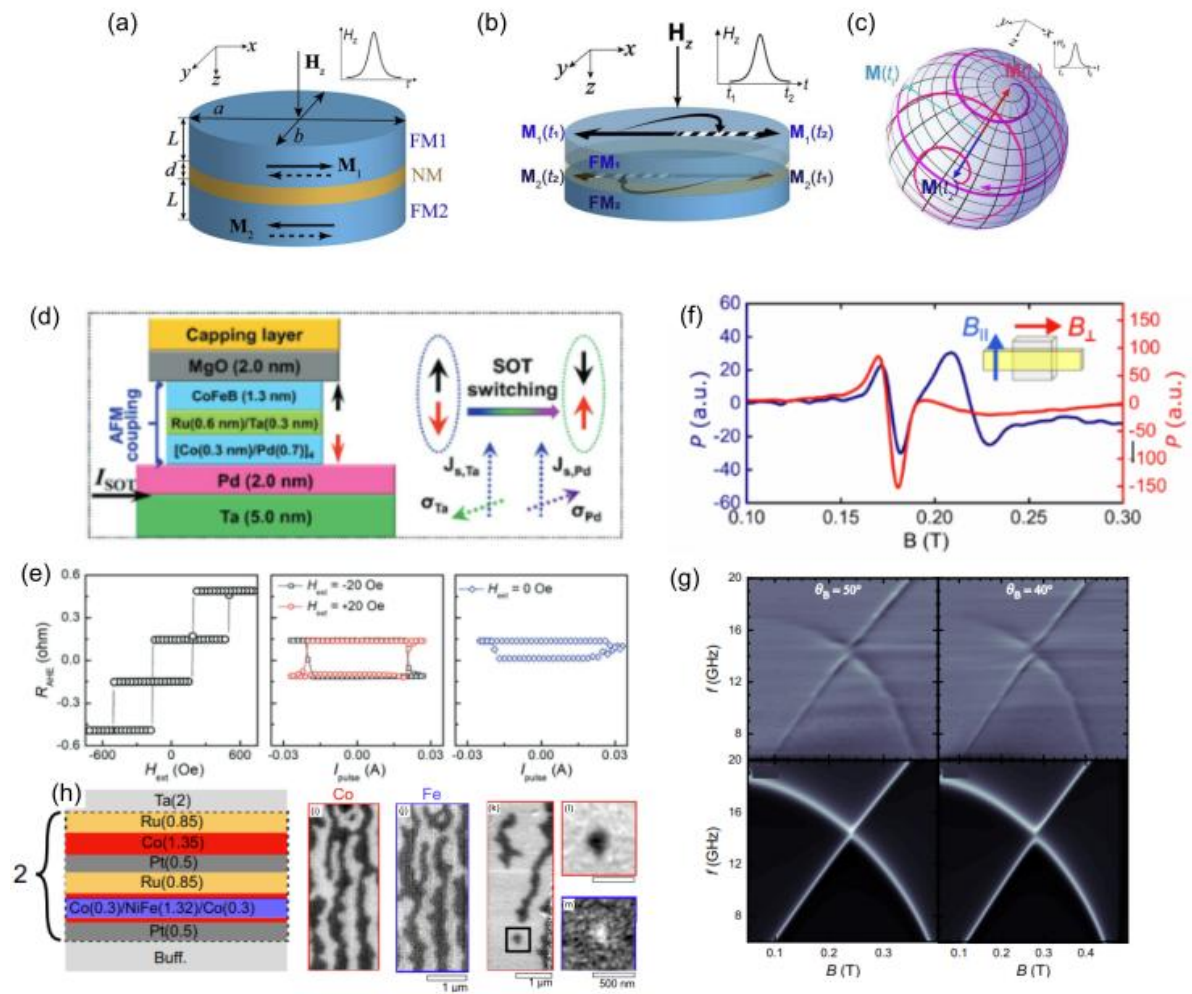

Fig 4. (a) **Schematic of SAF cell:** Two ferromagnetic layers (M1, M2) separated by a nonmagnetic spacer are switched via an out-of-plane magnetic pulse or circularly polarized laser. (b) **Barrier-free switching in SAF:** Rapid in-phase rotation enables ultrafast switching (b), unlike conventional ferromagnetic switching with precessional relaxation (c). (b-c are reprinted from [120]). (d) SOT switching in SyAF layer using a Ta/Pd bilayered spin Hall channel, enabling partial switching at zero magnetic field. (e) Anomalous Hall resistance ($R_{AHE}$) vs. $H_{ext}$ and $I_{pulse}$ for SOT switching in a 12-μm × 144-μm Hall bar device. Right panel is showing partial switching at zero field via Ta/Pd spin Hall channel. (d and e are adapted from [120]). (f) Magnetisation

dynamics studied using ferromagnetic resonance in SyAF. (g) A hybridization gap symbolic of magnon coupling is visible when the field is applied in out of plane direction and increases as the angle is changed. (f, g is reprinted from [121]). (h) **Observation of skyrmions in SyAF**: (h) Stack structure of SyAF. XMCD-STXM images at (i) Co $L_3$ and (j) Fe $L_3$ edges after out-of-plane demagnetization, showing isolated skyrmions at (k) top of image. XMCD microscopy image from ptychography reconstructions highlight skyrmion structure at (l) Co and (m) Fe edges. (h-l are adapted from [123]).

SyAFs operate in the gigahertz (GHz) frequency range, compared to the terahertz (THz) dynamics of natural antiferromagnets. This frequency range aligns with current spintronic technologies, such as MTJs in MRAM [119], where SyAFs stabilize magnetic configurations for efficient data storage and retrieval. In 2022, Zink *et al.* [119] demonstrated ultralow current switching of SyAF magnetic tunnel junctions using electric-field-assisted spin–orbit torque, as shown in Figure 4 (d,e). Recently, Dzhezherya *et al.* [120] theoretically demonstrated barrier-free switching in SyAF nanoparticles, achieved under a short, small amplitude magnetic field pulse applied perpendicular to the plane, as illustrated in Figure 4 (b-c).

Additionally, the dynamic properties of SyAFs enable low-dissipation spin-wave propagation and tunable coupling, as demonstrated by Sud *et al*.in (2020) (see Figure 4 (f-g)), which is crucial for magnonic devices [121,122]. In 2022 a study by Juge *et al*. [123] demonstrated the stabilization of room-temperature antiferromagnetic skyrmions in SyAFs by engineering interlayer exchange coupling and chiral interactions. Their work revealed that isolated antiferromagnetic skyrmions can be controlled electrically, overcoming the limitations of ferromagnetic skyrmions, such as dipolar field-induced deflections. The findings, shown in Figure 4 (h-l), highlight the potential of SAF-based skyrmions for energy-efficient spintronic applications.

The tunable nature of SyAFs opens pathways for novel functionalities, including superfluid-like spin transport [124]. By partially cancelling dipolar fields, these materials enable spin currents to flow with minimal energy loss, paving the way for low-power, high-efficiency devices. Advances in thin-film deposition techniques, such as sputtering and molecular beam epitaxy (*discussed in the next section*), ensure the precise fabrication of SyAFs.

In summary, SyAFs serve as a bridge between ferromagnetic and antiferromagnetic materials, offering the advantages of both systems with unmatched flexibility and scalability. Their engineered properties make them essential for the future of spintronic device technology. Together, collinear, noncollinear, and synthetic antiferromagnetic materials form the backbone of modern spintronics.

Collinear AFMs provide stability and efficiency for memory technologies, noncollinear AFMs enable exotic quantum functionalities and ultrafast dynamics, and SyAFs combine the benefits of natural and engineered systems with tunable properties and scalability. Collectively, these materials pave the way for next-generation devices, fostering ultrafast operations, enhanced scalability, and new computing paradigms.

**Ferrimagnetic and Multiferroic Materials**

Ferrimagnetic and multiferroic materials are crucial to modern spintronics due to their unique properties and wide range of applications [125,126]. Ferrimagnets, characterized by antiparallel magnetic moments on sublattices of unequal magnitude, exhibit a compensation point where the net magnetization cancels out. This compensation results in ultrafast magnetization dynamics [127] and reduced stray fields, making ferrimagnets a suitable candidate for spintronic devices [128]. Recent research has highlighted the advantages of compensated ferrimagnets, which combine key characteristics of both ferromagnets and antiferromagnets. For example, rare-earth transition-metal (RE-TM) alloys like GdFeCo [129] and Mn-based materials such as $Mn_4N$ and $Mn_3Ga$, show enhanced spin dynamics with minimal energy dissipation, making them promising for high-speed racetrack memory and STT devices [130].

Unlike traditional ferromagnets, the low damping and tunable net magnetization in ferrimagnets enable efficient current-driven domain wall motion and skyrmion dynamics, which are essential for next-generation memory and logic applications [131-133].

Moreover, ferrimagnets enable all-optical switching (AOS), as demonstrated in 2011 by Radu *et al* in GdFeCo [134], where femtosecond laser pulses trigger ultrafast magnetization reversal, a phenomenon later confirmed in other studies [135, 136]. The tunable angular momentum compensation point ($T_A$) offers a unique platform for achieving Walker-breakdown-free domain wall motion and reduced skyrmion Hall effects, making them a promising material class for ultrafast and energy-efficient spintronic devices [131, 132, 137]. Ferrimagnetic insulators like yttrium iron garnet (YIG), have also played a crucial role in spin-wave-based computing. In 2020, Collet *et al*. [138] showed how YIG thin films enable low-damping magnon propagation, facilitating energy-efficient spin transport for information processing.

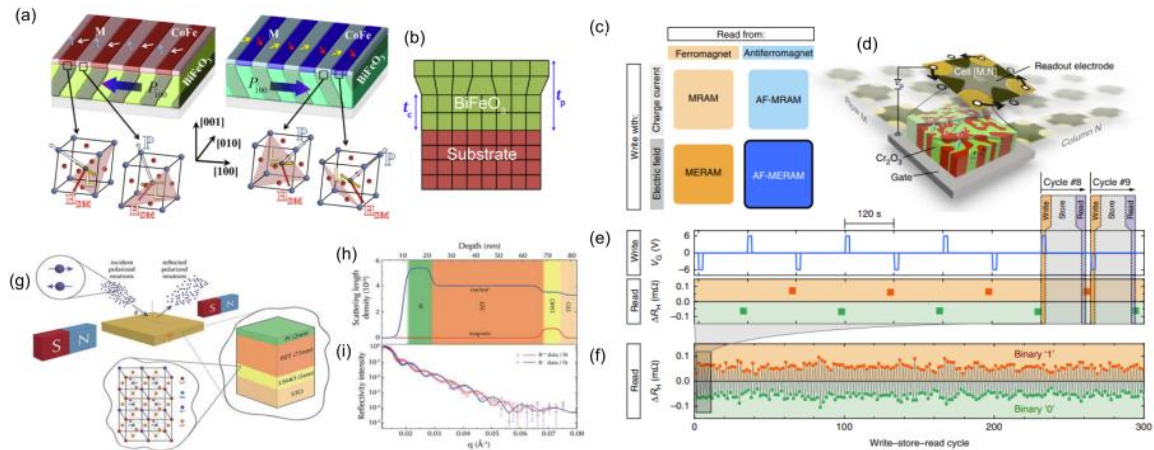

Fig 5. **Voltage-controlled magnetism in CoFe/BiFeO₃ (BFO) heterostructure**: (a) Phase-field model of coupled magnetic and ferroelectric domains in CoFe/BFO, showing electric-field switching of the interfacial exchange coupling field (HDM) perpendicular to electric polarization (P) and the antiferromagnetic axis (L) in BFO. Arrows represent local magnetization directions. (b) Schematic of an epitaxial BFO film with relaxed substrate clamping when the thickness exceeds the critical value ($t_c$) for interfacial dislocation generation. Reprinted from [141]. **Electric field-driven manipulation of the antiferromagnetic order parameter**. (c) AF-MERAM, a new antiferromagnetic spintronics memory concept. (d) Sketch of a memory cell in a device matrix with contact permutation for Hall readings. (e) Memory operation using voltage pulses to write binary information to the antiferromagnetic order parameter. ((c-f) are reprinted from Kosub *et al*. [142]) (f) Device

performance over 300 write–store–read cycles. (g) Schematic of polarized neutron reflectometry (PNR) experiment showing the sample with Pt and Nb-STO electrodes, and magnetization measurements using non-spin flip reflectivity. (h) Nuclear and magnetic scattering depth profiles, with enhanced magnetic scattering in the LSMO film. (i) Reflectivity intensities ($R^{++}$ and $R^{--}$) vs. momentum transfer vector $Q_z$ at room temperature with a 1 T in-plane field. Adapted from [144].

Multiferroic materials, which exhibit both ferroelectric and magnetic ordering, are highly promising for spintronics applications [139]. Bismuth ferrite (BiFeO$_3$), a well-known multiferroic, demonstrates room-temperature ferroelectricity and weak antiferromagnetic order, enabling electric-field-controlled magnetization switching. In 2007, Lebeugle *et al*. [140] demonstrated robust ferroelectricity in BiFeO$_3$ thin films, highlighting its potential for memory applications. In 2014 Wang *et al*. [141] used phase-field simulations (see Figure 5 (a-b).) to show that strain, combined with exchange interaction, can drive pure voltage-controlled 180° magnetization reversal in magnetic/BiFeO$_3$ heterostructures. This effect is attributed to the ferroelastic properties of BiFeO$_3$, which allow voltage-induced strain to modulate magnetism through magnetoelastic coupling. The electric-field manipulation of magnetization is crucial for low-power spintronic memory and logic devices. A significant development in this area occurred in 2016 when Kosub *et al*. [142] introduced purely antiferromagnetic magnetoelectric RAM (AF-MERAM) using Cr$_2$O$_3$. This design eliminates ferromagnetic components and relies on voltage-driven manipulation of the Néel vector, reducing the writing threshold by a factor of 50 as shown in Figure 5 (c-f). Similarly, in 2018, Shiratsuchi *et al*. [143] demonstrated exchange bias reversal in Cr$_2$O$_3$ heterostructures, further advancing voltage-controlled spintronics. In 2019, Paudel *et al*. [144] investigated the magnetoelectric coupling at the interface of a ferroelectric PbZr$_{0.2}$Ti$_{0.8}$O$_3$ (PZT) and magnetic La$_{0.67}$Sr$_{0.33}$MnO$_3$ (LSMO) heterostructure. Using polarized neutron reflectometry (PNR) and density functional theory (DFT) calculations (Figure 5 (g-l)), they studied how external electric fields influence the interfacial magnetic phases. Their findings revealed that switching the ferroelectric polarization direction of the PZT layer modulates the interfacial magnetic ordering in the LSMO layer, transitioning between ferromagnetic and A-type antiferromagnetic phases. This modulation is attributed to a combination of strain and charge-mediated effects, emphasizing the complex interplay between electric and magnetic properties in these heterostructures. By integrating magnetoelectric coupling in multiferroic materials, novel spintronic devices can achieve fast, low-power, and non-volatile operation, laying the foundation for next-generation memory and computing technologies.

While ferrimagnets offer ultrafast dynamics and multiferroics provide multifunctionality, both face challenges. Multiferroics require stronger coupling strength between electric and magnetic orders, while ferrimagnets need improved thermal stability. Advances in material synthesis, such as thin-film deposition techniques, are critical for overcoming these obstacles and realizing the full potential in spintronic, magnonic, and neuromorphic technologies.

**1.3 Fabrication and Characterization Techniques**

Fabrication and characterization are pivotal to the advancement of spintronic materials, ensuring precise control over their morphological, structural, electronic, and magnetic properties. These techniques bridge theoretical concepts with practical applications, enabling the design of materials tailored for specific functionalities. With continuous advancements in both fabrication and characterization technologies, researchers are now able to manipulate materials at the atomic and nanoscale levels, gaining unprecedented insights into their behavior. This synergy drives innovation in device performance and reliability, pushing the boundaries of memory, logic, and sensor applications. Below, we briefly describe the various fabrication and characterization techniques used in spintronic materials.

**Bulk and Single-Crystal Formation**

Bulk and single-crystal growth techniques form the foundation for understanding the intrinsic properties of materials. Methods such as the Czochralski process [145], where a seed crystal is pulled from molten material, facilitate the growth of uniform, defect-free crystals crucial for electronic and spintronic applications. The Bridgman technique [145], which employs controlled cooling within an ampoule, ensures high crystallographic quality and uniformity. For complex oxides like BiFeO$_3$ and Mn-based antiferromagnets, the floating zone (FZ) method [146] eliminates contamination risks by avoiding the use of crucibles, achieving high purity and precise stoichiometry. These techniques yield large, high-quality crystals that are indispensable for both fundamental studies and device integration.

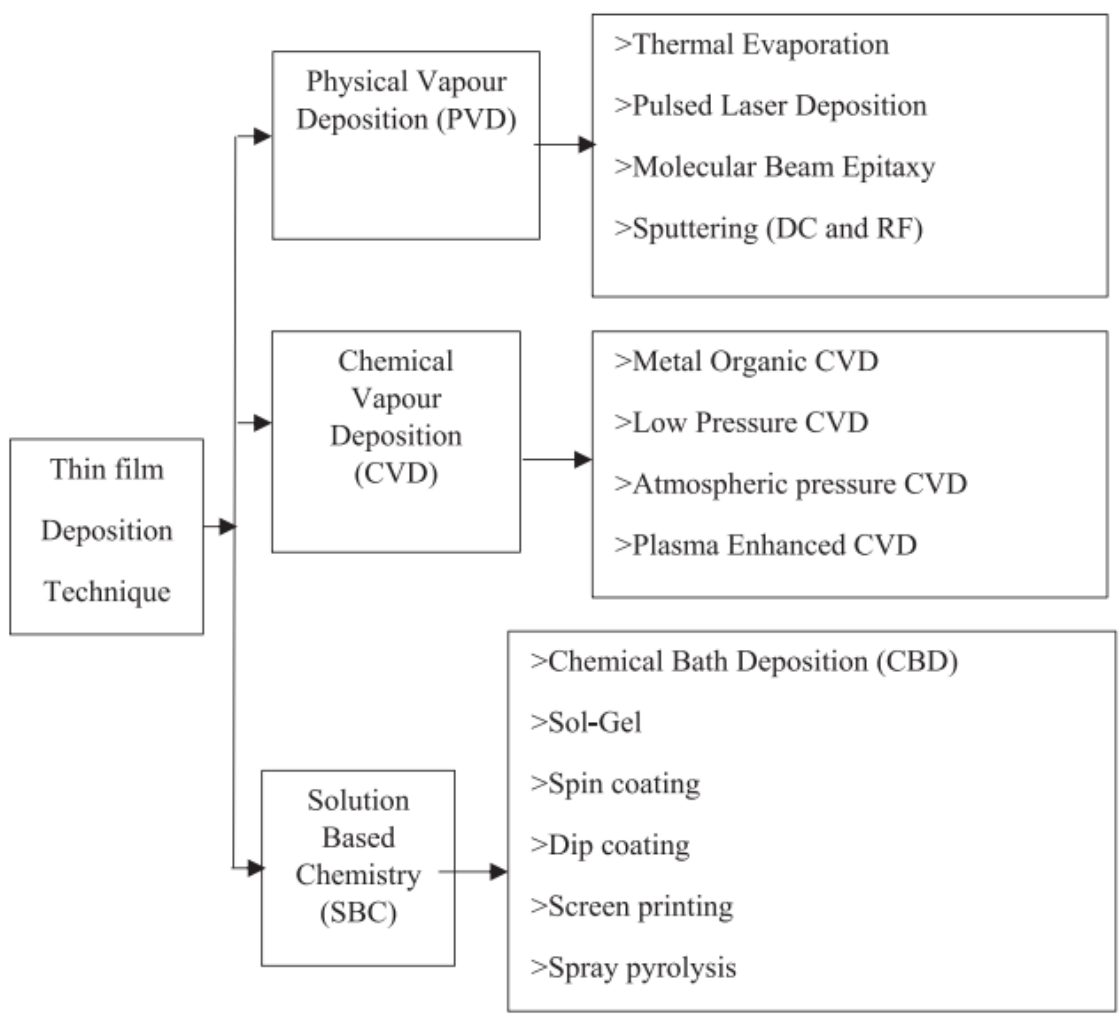

Fig 6. Classification of different thin film deposition methods. Reprinted from [147].

**Thin Film Deposition**

Thin film deposition techniques enable the fabrication of nanoscale materials with precise control over thickness, composition, and uniformity—key parameters for spintronics. Figure 6 [147] illustrates different methods for thin film deposition. Physical Vapor Deposition (PVD) methods, such as sputtering and Molecular Beam Epitaxy (MBE), produce defect-free films under vacuum conditions, making them ideal for magnetic multilayers. Chemical approaches like Chemical Vapor Deposition (CVD) and Plasma-Enhanced CVD (PECVD) utilize gaseous precursors for conformal coatings, while Pulsed Laser Deposition (PLD) employs laser ablation to achieve stoichiometrically accurate films. The precision of Atomic Layer Deposition (ALD), which deposits films one atomic layer at a time, is essential for applications requiring atomic-scale control. Cost-effective solution-based methods like sol-gel deposition are particularly suited for oxide thin films. Together, these techniques enable the creation of materials optimized for memory, sensors, and energy-efficient devices, driving the advancement of advanced spintronic technologies. For a more detailed study of these deposition techniques, the readers can refer to Oke *et al.* [148].

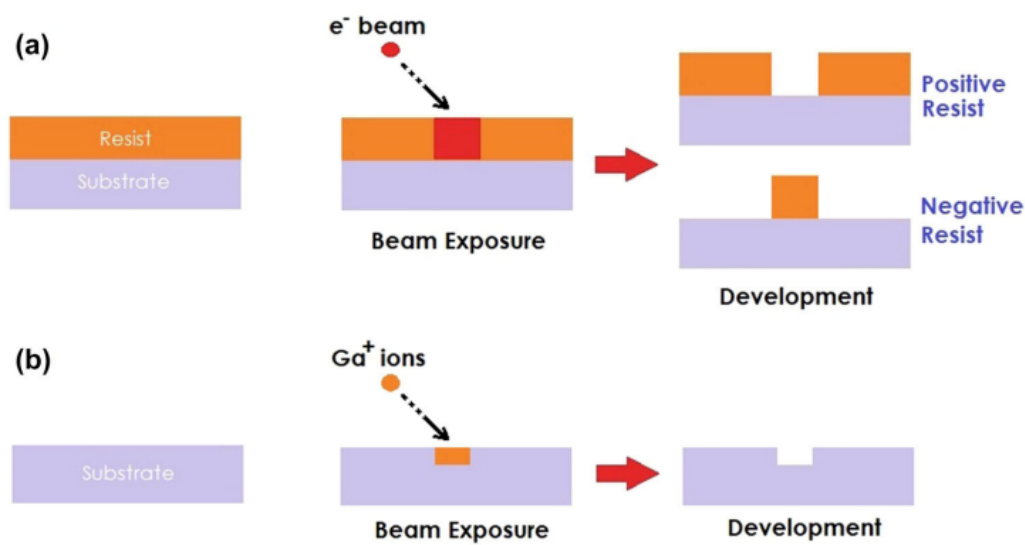

Fig 7. Fabrication process for electron beam lithography and (b) Focussed ion beam lithography. Reprinted from [149].

**Nanostructuring Techniques**

Nanostructuring methods are crucial for fabricating devices at the nanoscale, providing precision in patterning and enhancing functionality. Electron Beam Lithography (EBL) [149], which offers sub-10 nm resolution, is widely used for quantum and spintronic devices, while optical lithography remains the backbone of semiconductor manufacturing, achieving resolutions in the tens of micrometers. Focused Ion Beam (FIB) processing [150] enables direct material etching, milling, and deposition, making it ideal for creating intricate 3D nanostructures and preparing high-quality TEM samples. Figure 7 demonstrates the EBL and FIB nanostructuring techniques. Nanoimprint Lithography (NIL) [151] offers cost-effective, high-throughput patterning, while Atomic Force Microscopy (AFM) lithography [152] enables surface modifications at the atomic scale. Emerging techniques, such as Direct-Write Laser Lithography [153] and Two-Photon Lithography [154], allow for rapid prototyping and the fabrication of complex 3D structures. Bottom-up approaches, including self-assembly and chemical patterning [154], further expand the toolkit for nanostructure fabrication, enabling innovations in miniaturized device design. For a more detailed review, refer to [154].

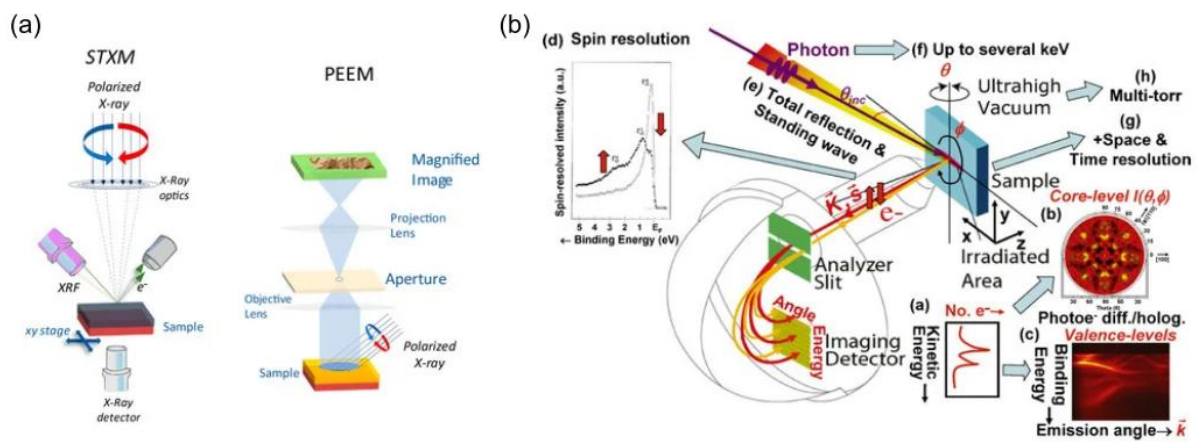

Fig 8. (a) XMCD based microscopy techniques. (b) Illustration of different information obtained from XPS technique. Reprinted from [156].

**Characterization Techniques**

Understanding the macroscopic magnetic properties of thin films and multilayers requires a systemic approach that combines complementary characterization technique. Imaging methods such as Scanning Electron Microscopy (SEM) and Transmission Electron Microscopy (TEM) [155] provide nanoscale insights into morphology and internal structures.

Advanced variants, like High-Resolution TEM (HRTEM) and Scanning Transmission Electron Microscopy (STEM) [156], offer atomic-level resolution. Magnetic domain imaging [157] can be achieved through techniques like Spin-Polarized Low-Energy Electron Microscopy (SPLEEM) and Photoemission Electron Microscopy (PEEM). For structural analysis [158], X-ray Reflectivity (XRR), High-Resolution X-ray Diffraction (HRXRD), and Grazing Incidence X-ray Diffraction (GIXRD) are used to explore interfacial roughness, crystallinity, and strain. Techniques like Polarized Neutron Reflectometry (PNR) further elucidate buried magnetic profiles and layer interfaces. Spectroscopic techniques such as X-ray Magnetic Circular Dichroism (XMCD) and X-ray Photoelectron Spectroscopy (XPS) [159] analyze chemical states and magnetic properties at the element-specific level. The experimental setup is shown in Figure 8. Resonant Inelastic X-ray Scattering (RIXS) and X-ray Fluorescence (XRF) provide valuable insights into electronic excitations and elemental compositions. Magnetization dynamics are studied using Ferromagnetic Resonance (FMR), Brillouin Light Scattering (BLS), and Time-Resolved Magneto-Optical Kerr Effect (TR-MOKE) [160], offering a comprehensive understanding of spin precession, damping, and ultrafast spin dynamics.

Characterization of SOT efficiency is crucial for developing energy-efficient spintronic devices. Techniques like Harmonic Hall Effect and Spin-Torque Ferromagnetic Resonance (ST-FMR) [161] measure SOT contributions and spin current generation, while magnetization switching experiments assess field-free switching and critical current densities, providing essential optimization of materials for next-generation applications.

**1.4 Current and Emerging Research in Magnetic Materials for Spintronics**

Current spintronics research prioritizes low-power, high-efficiency devices that harness nanoscale magnetic phenomena to meet the growing demand for energy-efficient technologies. Emerging trends, such as magnonics, SOT materials, phonon-magnon coupling, and 3D spin textures, are reshaping the field by enhancing spin dynamics, enabling robust data storage, and introducing novel functionalities [162,163]. The interplay of geometry, topology, and intrinsic material properties plays a key role in these advancements [163]. 3D spin textures allow manipulation of magnetic states beyond planar systems, while magnonics [164] and phonon-magnon interactions enable efficient signal processing and quantum computing. SOT materials [165] are foundational next-generation non-volatile memory and logic devices. These breakthroughs are driven by advanced fabrication tools, which support the study of spin frustration, noncollinear spin textures, and dynamic interactions. This section highlights these developments and their transformative potential in spintronics.

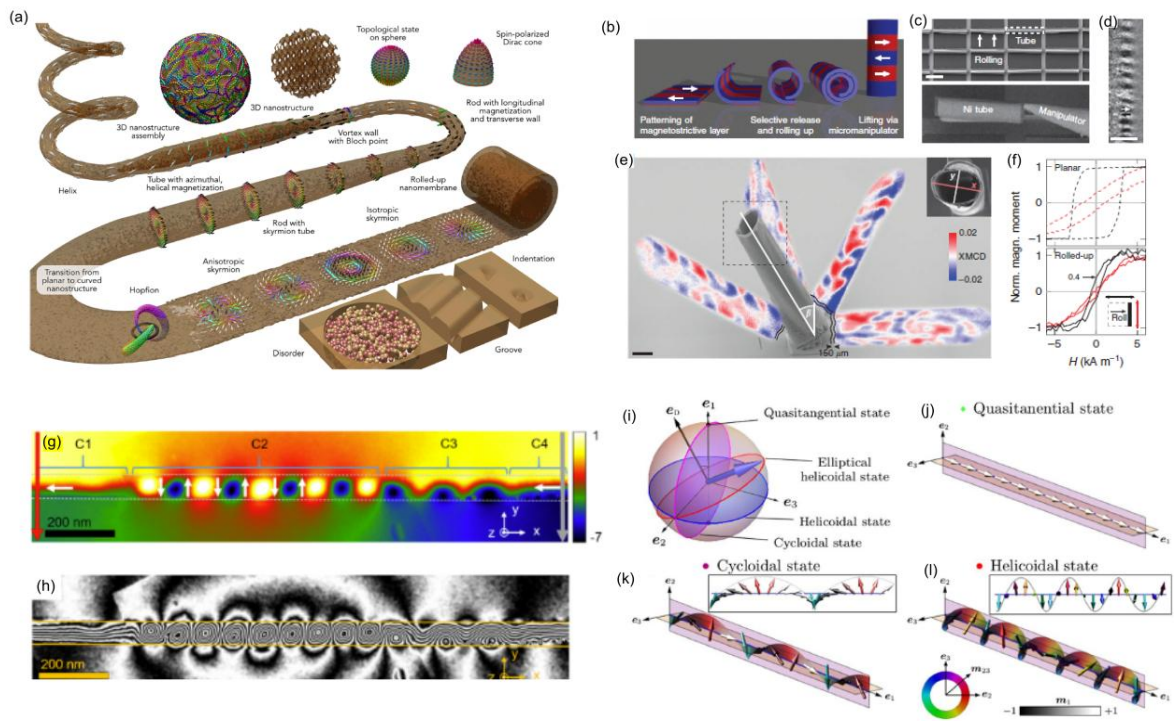

Fig 9. (a) Magnetism in Curved Geometries Across Real, Reciprocal, and Spin Space. Curvature, strain, and short-range order break local inversion symmetry, enabling 3D topological spin textures via emergent DMI. Reprinted from [169]. **Fabrication and characterization of circulating spin textures:** (b) Azimuthally magnetized tubular architectures are created by rolling up prepatterned strained nanomembranes. (c) Electron micrographs show the structure of these rolled-up tubes. (d) Kerr microscopy visualizes the transverse magnetization at the tube's top. (e) XMCD shadow contrast patterns obtained via T-XPEEM highlight the circulating magnetization in a vertically fixed Ni tube. (f) Magnetic hysteresis loops confirm the presence of azimuthal or helical magnetization within the rolled-up nanomembranes. (b-f are reprinted from [167]). (g) Unwrapped magnetic phase shift image of a $Co_{85}Ni_{15}$ nanowire (70 nm diameter), obtained using electron holography, with phase shift in radians represented by a color scale. (h) Magnetic flux lines extracted from (g) using a cosine transformation, amplified by a factor of 4. (g-h) are adapted from [171]. **Phase diagrams of equilibrium magnetization state in helical wires**: (i) Magnetization vector evolution on a unit sphere for different states. (j–l) Schematics of equilibrium states for varying mesoscale DMI, including quasitangential,

cycloidal and helicoidal states. (j-l) are reprinted from [172].

**3D Spin Textures: Properties, Materials, and Advanced Applications**

3D spin textures represent a significant breakthrough in magnetism and spintronics, characterized by complex, non-collinear magnetization vector fields in three dimensions. Unlike their 2D counterparts, these textures arise from intricate interactions such as spin frustration, Dzyaloshinskii-Moriya Interaction (DMI), and symmetry-breaking effects in curved or disordered systems [167, 169], as illustrated in Figure 9(a). The development and control of 3D spin textures are enabled by advances in manipulating magnetization through curvature, topology, and tailored material design.

Materials such as synthetic multilayers (e.g., Pt/Co/Ta), transition metal oxides, and disordered ferrimagnets provide fertile ground for stabilizing 3D textures [168]. The interplay of strain, curvature, and local DMI enables the creation of complex magnetic textures, such as hopfions and skyrmioniums, which are unattainable in planar systems. Chen *et al*. (2022) [168] investigated the evolution of chiral spin textures in Co/Pt-based multilayers, demonstrating how increasing chiral interactions influence domain wall helicity, domain compressibility, and skyrmion formation. By combining advanced microscopy and simulations, their work established a microscopic framework for tailoring spin textures in multilayer films. In 2015 Streubel *et al.* [167] introduced a method to visualize and reconstruct the 3D magnetic domain structures of curved tubular magnetic Ni thin films using soft X-ray microscopy, as shown in Figure 9 (b-f). The 3D magnetization arrangement was reconstructed by analyzing changes in magnetic contrast across multiple projections at varying angles, offering detailed insights into domain patterns and magnetic interactions within the structure. This approach significantly advanced the study of 3D magnetization in complex-shaped thin films. The coupling of magnetization with electronic effects, such as the topological Hall effect [170] and SOT, enables efficient manipulation and detection of magnetic states. In 2019, Andersen *et al.* [171] investigated 3D nanostructured Co-rich CoNi nanowires, revealing a correlation between local structural, chemical properties and magnetization configurations. As depicted in Figure 9 (g-h), their work utilized electron holography to demonstrate how grain boundaries and engineered segmentation transform axial magnetization into vortex or transverse states. These findings are critical for spin torque nano-oscillators and high-velocity domain wall motion in spintronic applications. Similarly, Volkov *et al.* [172] introduced a mesoscale DMI that combines intrinsic spin-orbit and curvature-driven effects to create tunable chiral magnetic textures (Figure 9 (i-l)). These properties position 3D spin textures as ideal candidates for energy-efficient and high-density devices, such as racetrack memory and neuromorphic computing.

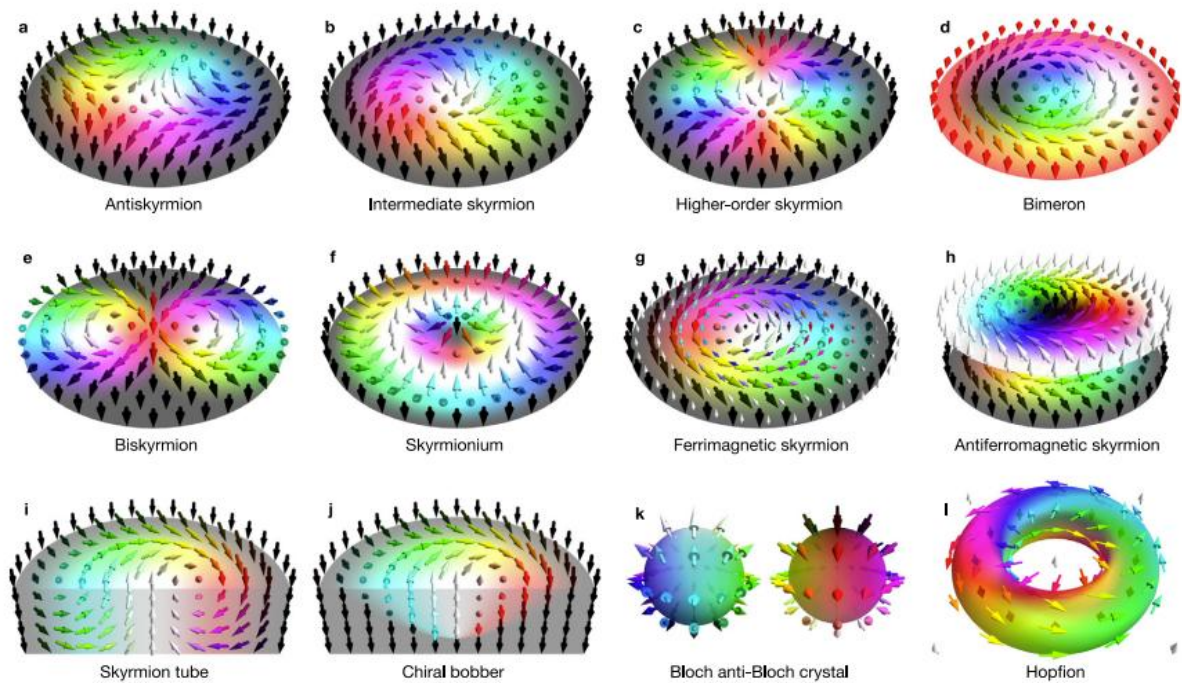

Fig 10. **Overview of topologically non-trivial skyrmion spin textures.** (a–c) Various skyrmions with different helicities and vorticities, including antiskyrmions, intermediate-helicity skyrmions, and higher-order skyrmions. (d) Magnetic bimeron as a skyrmionic excitation in an in-plane magnetized medium. (e–h) Skyrmion pairs, including biskyrmions, skyrmioniums, and ferrimagnetic/synthetic antiferromagnetic skyrmions. (i–l) 3D extensions such as skyrmion tubes, chiral bobbers, hedgehog lattices, and hopfions. Colored arrows indicate magnetic moment orientations. Reprinted from [173].

**Domain Walls and Skyrmions**

Domain walls and skyrmions [173,174] are key examples of 3D spin textures that display unique behaviors in curved geometries and advanced materials. In 3D structures such as curved nanowires and tubular geometries [175], domain walls are significantly influenced by local curvature and anisotropies, resulting in novel dynamics. For instance, NiFe nanotubes exhibit curvature-dependent magnetization reversal processes, enabling precise control over domain wall motion [175]. Studies involving experiments and simulations have demonstrated that, in thicker films, domain wall spin textures diverge from symmetric Bloch or Néel configurations, presenting features like Néel caps in nanoelements or hybrid Néel-Bloch walls in ferrimagnets [176-178]. Additionally, nanowire switching has been shown to involve transient 3D spin textures, such as Bloch points and lines, which have been predicted by micromagnetic simulations and are now being actively explored in experimental investigations [179]. Skyrmions are another hallmark of 3D spin textures, first observed in MnSi using reciprocal space measurements [174]. Figure 10 illustrates various topologically non-trivial skyrmion spin textures with different types of helicities and vorticities. These nanoscale, topologically protected structures are stabilized by DMI and spin frustration in materials such as SyAFs [180] and oxide heterostructures. Numerous studies have demonstrated the room-temperature stabilization of antiferromagnetic skyrmions, underscoring their potential for spintronic applications. Zhang *et al*. (2016) [181] showed that SyAF multilayers with interfacial DMI can host stable skyrmion lattices at room temperature. Additionally, Correia *et al*. (2020) [182] confirmed the stabilization of antiferromagnetic skyrmions in synthetic multilayers through RKKY-mediated interlayer exchange, demonstrating their robustness against thermal fluctuations. In parallel, oxide heterostructures have emerged as another promising platform, where interfacial engineering for stabilization of antiferromagnetic skyrmions at room temperature through interfacial engineering. For example, Zhou *et al*. [183] observed antiferromagnetic skyrmions in Pt/Co/GdOx heterostructures, while Das *et al*. (2021) [184] demonstrated the stabilization of polar skyrmion bubbles in PbTiO$_3$/SrTiO$_3$-based heterostructures. These 3D skyrmion structures differ significantly from the 2D skyrmions observed in other magnetic systems. Legrand *et al*. (2020) [180] further reported the generation and tunability of antiferromagnetic skyrmions in synthetic antiferromagnets, highlighting their efficient manipulation for device applications. Collectively, these studies establish synthetic antiferromagnets and oxide heterostructures as promising platforms for advancing room-temperature skyrmion-based spintronic technologies The ability to stabilize room-temperature skyrmions in multilayer stacks and chemically disordered materials has significantly expanded their applicability. Furthermore, the exploration of skyrmions in 3D systems has revealed new spin textures, such as hopfions—complex toroidal configurations with intertwined topologies [185]. These higher-order textures exhibit unique dynamical properties, including suppressed gyroscopic effects compared to skyrmions, leading to more stable and predictable motion. This makes them promising candidates for high-speed and energy-efficient data transport in spintronic circuits, as they circumvent the skyrmion Hall effect that typically causes lateral drift. Due to their compact size, stability, and low-energy manipulation, skyrmions are particularly well suited for racetrack memory applications. In thin-film multilayers [188], for instance, skyrmions have been shown to move efficiently under low-current densities, enabling high-density and reliable data storage. Materials such as ferrimagnets and 3D nanoparticle assemblies are crucial for investigating domain wall and skyrmion behaviors [186]. These systems enable the tuning of spin textures through local variations in exchange interactions and DMI, providing new pathways for energy-efficient data processing. Integrating 3D spin textures into spintronic devices holds significant promise for advancing memory, logic, and computing technologies. Additionally, curved nanostructures, such as helical nanowires and toroidal geometries, enhance spin-wave propagation and enable the creation of tunable magnonic crystals, further expanding the role of 3D spin textures in neuromorphic architectures.

**Topological Phases and 3D Structures**

The emergence of 3D spin textures is intrinsically linked to topological phases and the interplay of curvature, frustration, and DMI [169]. Amorphous ferrimagnets [189], disordered alloys [190], and chiral crystals [191] serve as versatile platforms for realizing these phases, often circumventing the need for large DMI. For instance, disordered Co-Si alloys can stabilize nanoscale skyrmions through DMI and spin frustration [192]. Chemical and structural disorders can induce bulk DMI and stabilize topological states by enhancing Anderson localization [193,194], which suppresses electron transfer while increasing the local density of states. This, in turn, enhances spin–orbit coupling, local DMI, magnetoresistance, and Hall effects [195].

Curvature plays a pivotal role in the stabilization and manipulation of these textures. Tubular geometries and 3D networks in permalloy nanostructures or oxide-based materials enable dynamic control of spin textures through strain, temperature, or external fields [169]. These capabilities are vital for the development of reconfigurable magnonic devices and 3D spintronic circuits, where local modulation of magnetic exchange interactions is key.

The ability to harness 3D topological states for device applications marks a significant advancement in spintronics. Notable examples include 3D magnonic crystals [196], which can be tuned via curvature and DMI, and chiral spin textures that enhance spin-wave selectivity. These innovations pave the way for the creation of tunable, low-power systems suited for next-generation computing and signal processing. The exploration of 3D spin textures has unveiled a new paradigm in spintronics, bridging fundamental physics with practical applications. Materials such as synthetic multilayers, ferrimagnets, and curved nanostructures provide platforms for stabilizing complex textures like domain walls, skyrmions, and hopfions. Their unique properties, combined with advanced manipulation techniques, offer immense potential for applications ranging from memory devices to neuromorphic computing and magnonics. As advancements in synthesis, modeling, and characterization continue, 3D spin textures are set to revolutionize energy-efficient and high-performance spintronic technologies.

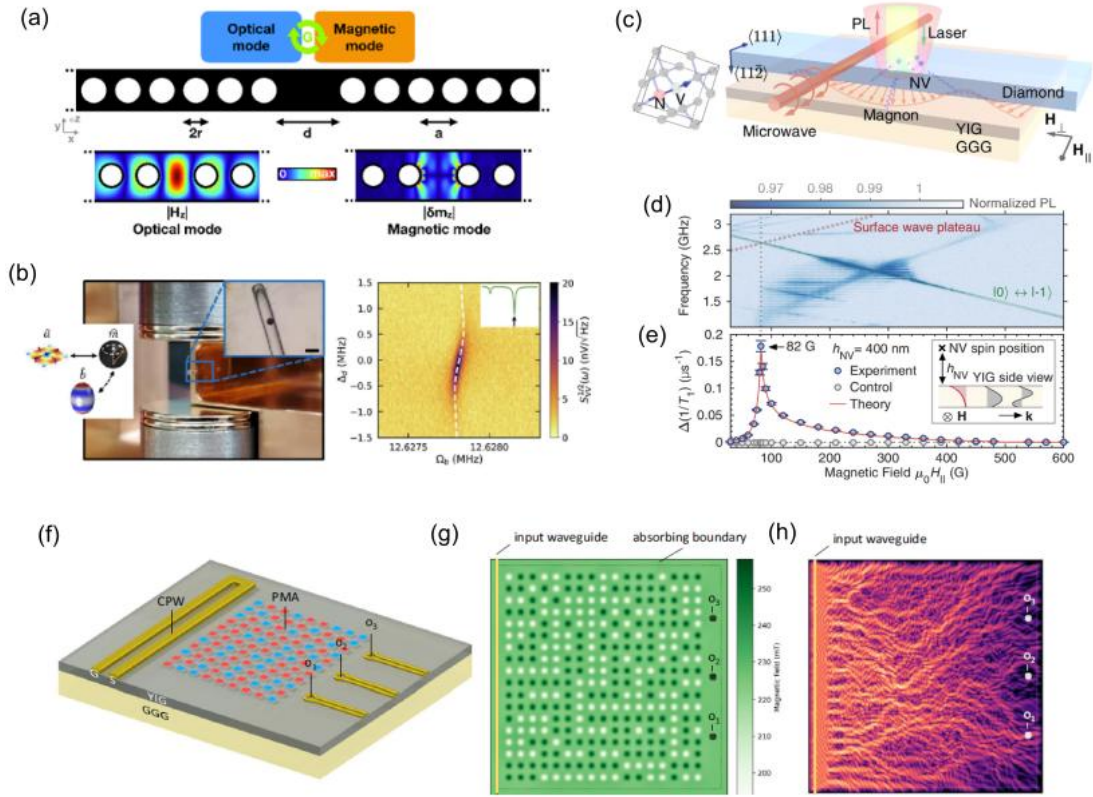

Fig 11. (a) Design of a 1D optomagnonic crystal using a magnetic dielectric (YIG). A hole array with a central defect act as a Bragg mirror, co-localizing optical and magnon modes, enabling enhanced cooperativity. Adapted from [212] (b) Prototype cavity magnon-mechanical system. (Left) A YIG sphere in a microwave cavity couples its Kittel mode resonantly to the cavity and off-resonantly to a mechanical mode. (Right) Tuning to triple resonance enables control of dynamical backaction effects like the magnon spring effect. Adapted from [213]. (c) Schematic of NV center spins coupling with magnons. **Surface-magnon induced relaxation.** (d) ODMR of NV centers on YIG, showing NV transition (green) and magnon plateau (red). (e) NV relaxation rate with experiment (blue/gray) and theory (red). Inset: YIG cross-section with magnon modes. Adapted from [214]. **Nanomagnet-based non-linear spin-wave scatterer.** (f) Computing device schematic where CPW input signals and programming magnets on YIG define weights. (g) Magnets with perpendicular anisotropy create a bias-field landscape, optimized via training. (h) Spin-wave intensity pattern showing high output in a 10 μm × 10 μm simulation area. (f-h) are reprinted from [216].

**Magnonics: Current and Emerging Materials for Advanced Applications**

Magnonics, focused on the study of spin waves (magnons) as quantized excitations in magnetically ordered systems, represents a transformative field within spintronics [197].

Magnons facilitate energy-efficient information transport by carrying spin angular momentum, making them vital for developing low-power, next-generation devices. Advances in materials science have propelled the field, enabling devices that operate at terahertz frequencies while maintaining robust functionality. The unique properties of magnons, coupled with the ability to manipulate their behaviour, have solidified magnonics as a cornerstone of modern spintronic research [197]. Traditional materials like yttrium iron garnet (YIG) [198] and nickel ferrite ($NiFe_2O_4$) [199] are well-established for their exceptional magnonic properties, including long spin-wave lifetimes and minimal damping. While YIG is known for its ultra-low damping, $NiFe_2O_4$ exhibits higher damping but offers strong magnetoelastic coupling, making it valuable for spintronic and spin-wave transport applications. However, the attention is now shifting towards emerging materials with unique characteristics and functionalities. Synthetic multilayers, such as Pt/Co/Ta and Pt/Co/Ir, leverage interfacial effects like SOTs and DMI to stabilize chiral spin textures and non-reciprocal spin waves [200,201]. These engineered heterostructures provide unparalleled control over spin-wave dynamics, paving the way for compact, energy-efficient magnonic devices. 2D magnetic materials [202,203], such as $CrI_3$ and $CrSe_2$, have gained prominence due to their atomically thin structures and integration capabilities with van der Waals (vdW) heterostructures. These materials enable highly tunable magnonic properties, making them ideal for hybrid magnonics [204], where magnons interact with photons and phonons at the nanoscale. Meanwhile, chiral and topological materials, including B20 alloys [205] like FeGe and MnSi, are being explored for their ability to stabilize topologically protected states, such as skyrmions and hopfions. These materials, known for their robust spin-wave propagation and resilience to structural imperfections, play a crucial role in developing dissipationless magnonic systems. Emerging research also highlights the potential of amorphous ferrimagnets and disordered alloys, such as GdFeCo [206] and CoSi [207], in supporting nonlinear magnonic phenomena. These materials not only sustain spin-wave propagation [208] but also introduce local spin frustration, which enables the stabilization of complex spin textures [209] and the exploration of novel effects like three-magnon interactions and soliton formation. The interplay between structural disorder and magnetic properties has opened new avenues for designing reconfigurable magnonic devices. The development of 3D magnonic architectures has introduced materials like permalloy nanotubes and helical nanowires, where curvature and local anisotropy enable new pathways for spin-wave manipulation [169]. In parallel, non-Hermitian systems [210], which incorporate gain and loss dynamics, offer unique phenomena such as unidirectional spin-wave propagation and exceptional points, significantly expanding the design space of magnonic circuits. Hybrid magnonic systems [211], which couple magnons with photons or phonons, benefit from high-quality oxide heterostructures and garnet films, providing the necessary properties for quantum applications including microwave-to-optical transduction and magnon-photon coupling. The coherent coupling of magnons to microwave and optical photons holds promise for quantum frequency transduction, with cooperativity being a key figure of merit. A major challenge is boosting cooperativity, particularly in the optical regime, due to inherently weak coupling and mode-matching difficulties. Proposed solutions [212] include photonic crystals (see Figure 11 (a)) in magnetic dielectrics, phonon coupling, and magnon self-hybridization in complex heterostructures to enhance transduction efficiency. In tri-partite hybrid systems (Figure 11 (b)), magnons control phonon frequency and dissipation via dynamical effects [213]. Recent studies have explored Nitrogen-vacancy center (NV) -magnon coupling for quantum spintronics. Fukami *et al* [214] experimentally characterized magnon-mediated NV–NV interactions, advancing hybrid quantum architectures, as shown in Figure 11 (c-e). Nonlinear magnonics [215] is another rapidly evolving area, driven by materials like ferrimagnets and SyAFs. These systems support high spin-wave amplitudes and nonlinear effects, such as parametric amplification and magnon-magnon coupling. Such effects are crucial for enhancing weak spin signals and enabling frequency conversion, offering significant potential for magnon-based communication technologies. Papp *et al.* [216] introduced a spin-wave-based neural network hardware, where signal processing, routing, and activation occur via spin-wave propagation and interference, with magnetic-field-controlled weights and nonlinear behaviour enhancing computational power as shown in Figure 11 (f-h). Materials that amplify spin waves while maintaining low energy dissipation are pivotal in advancing this field. The applications of these materials span a wide range, from racetrack memory to neuromorphic computing. By integrating fundamental physics with innovative material design, magnonics is poised to drive the next wave of technological advancements in spintronics, quantum information processing, and beyond.

**Spin-Orbit Torque Materials: Advances and Applications**

Spin-orbit torque (SOT) materials represent a revolutionary class in spintronics, enabling efficient magnetization switching through spin-orbit coupling mechanisms. Unlike conventional STT systems, SOT materials utilize spin currents generated from charge currents in non-magnetic layers, allowing for field-free, low-power control of magnetization [217]. Figure 12 (a-d) shows the spin orbit coupling and SOT phenomena. *For more detailed description refer to* [217]. Recent advancements in materials science have broadened the scope of SOT-based systems, emphasizing the integration of emerging materials to optimize performance and functionality. Beyond traditional heavy metals like Pt, Ta, and W, research has shifted toward unconventional materials with enhanced spin-orbit coupling and tunable properties. Topological insulators (TIs), such as $Bi_2Se_3$ and $Bi_2Te_3$, have garnered significant attention due to their surface states, which exhibit spin-momentum locking. The integration of TIs into device architectures has paved the way for ultralow power and high-speed spintronic applications [218,219]. Two-dimensional (2D) materials are another emerging frontier. Transition metal dichalcogenides (TMDs), such as $MoTe_2$ and $WSe_2$, offer intrinsic spin-orbit coupling and excellent compatibility with van der Waals heterostructures [220]. When paired with magnetic layers, 2D materials enable precise control over spin current direction and magnitude, opening up possibilities for flexible and transparent spintronic devices. Additionally, graphene-based heterostructures with proximity-induced spin-orbit coupling have shown potential for novel SOT mechanisms [221].

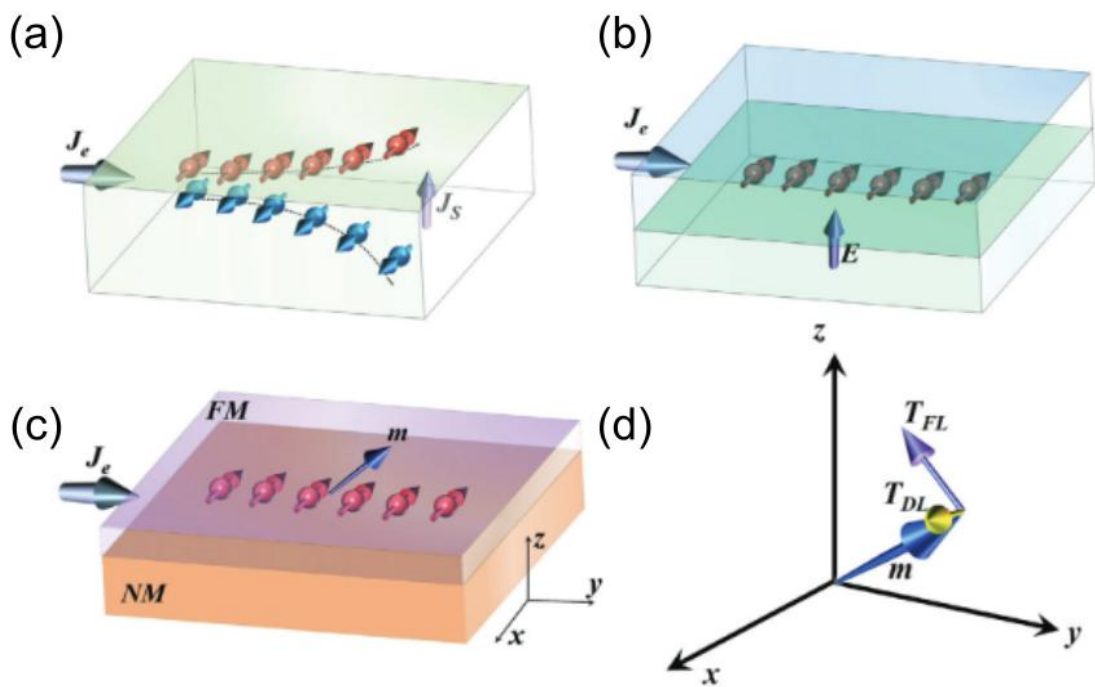

Fig 12. (a) **Illustration of SOC and SOT phenomena.** (a) Spin Hall effect schematic. (b) Rashba–Edelstein effect schematic. (c) Spin accumulation at the FM/NM interface in a heterostructure. (d) SOT schematic with field-like (purple) and damping-like (yellow) torques. Reprinted from [222].

Oxide-based systems, such as $SrIrO_3$ and $LaAlO_3/SrTiO_3$ interfaces, provide another promising platform. These materials exhibit tunable Rashba spin-orbit coupling and high thermal stability, making them suitable for integration into silicon-based technologies. Oxide heterostructures also enable for ionic control of magnetization, opening new possibilities for energy-efficient and reconfigurable devices. In 2019 Liu *et al.* [223] demonstrated all-oxide SOT devices using $SrIrO_3/SrRuO_3$ bilayers, achieving field-free magnetization switching and efficient charge-to-spin conversion, paving the way for low-power magnetic memory application. Non-Hermitian system, characterized by their incorporation of gain and loss mechanisms, have recently emerged as a novel class for SOT research [224, 225]. These systems enable non-reciprocal spin current generation and amplification of spin waves, which are crucial for advanced magnonic and spintronic circuits.

The practical applications of SOT materials are vast, spanning memory devices, logic and computing architectures. In magnetic memory, SOT materials facilitate reliable magnetization switching at lower current densities compared to STT systems. For example, Pt/Co/Ir stacks have demonstrated high endurance and scalability in SOT-MRAM [226,227]. The field-free switching enabled by TIs and heavy metal multilayers further enhances device performance, making them ideal for high-density storage solutions [228]. In neuromorphic computing, SOT materials facilitate the development of spintronic neurons and synapses [229]. SyAFs, in particular, show great promise in this domain due to their ability to exhibit ultrafast dynamics and minimal crosstalk between neighbouring elements.

Hybrid magnonic-SOT systems [164] represent an emerging trend, combining the low-dissipation properties of magnons with the high efficiency of spin-orbit torque. These systems leverage materials like YIG/Pt bilayers [230], where the SOT drives spin-wave excitation in the YIG layer. Such hybrid systems are poised to revolutionize quantum information processing by enabling coherent magnon manipulation and coupling with photons or phonons.

Looking ahead, integration of SOT materials into three-dimensional (3D) architectures is a critical research direction [231]. Additionally, the exploration of multifunctional materials, such as multiferroics, which couple electric and magnetic properties, holds great promise. These systems allow electric-field control of SOT, reducing power consumption and expanding the range of device applications. For example, $BiFeO_3$-based heterostructures allow for direct electric-field manipulation of magnetization via SOT, offering a new paradigm for energy-efficient spintronics [232].

The rapid progress in spin-orbit torque materials highlights s their transformative potential in spintronics. As materials science advances, SOT materials are set to become the cornerstone of next-generation spintronic devices, enabling energy-efficient and scalable solutions for modern technological challenges.

**1.5 Novel Spintronic applications**

As the demand for energy-efficient and high-performance devices grows, the exploration of innovative applications for magnetic materials continues to expand. These materials, with their unique magnetic properties and tunability, offer unprecedented opportunities to revolutionize both traditional and emerging technologies. This section explores how advanced magnetic materials are being utilized in memory devices, nano-oscillators, and unconventional computation platforms, showcasing their potential to redefine the landscape of modern electronics and information processing. By focusing on both established and cutting-edge applications, this section highlights the versatility and importance of magnetic materials in driving future technological breakthroughs.

**Materials for memory applications**

**Magnetic Tunnel Junctions (MTJs)**: Revolutionizing MRAM Technology- MTJs form the cornerstone of MRAM technology, leveraging the tunnelling magnetoresistance (TMR) effect. This effect arises from the dependence of electrical resistance on the relative magnetization alignment of ferromagnetic layers separated by a thin insulating barrier. Early implementations in the 1990s utilized amorphous $AlO_x$ barriers [233], achieving moderate TMR ratios of around 30% at room temperature. However, a breakthrough came with the introduction of crystalline MgO barriers in 2004, which enabled TMR ratios exceeding 200% at room temperature, as reported in seminal studies by Parkin *et al*. [234] and Yuasa *et al*. [235] MgO's unique coherent tunneling properties maximize spin polarization, significantly boosting device efficiency and performance. TMR ratios exceeding 1000% have since been reached [236].

The storage layer in MTJs typically employs CoFeB, a material celebrated for its high spin polarization and excellent interface properties with MgO. In 2010, Ikeda *et al.* [237] demonstrated that CoFeB/MgO interfaces exhibit robust thermal stability, significantly advancing the development of reliable magnetic random-access memory (MRAM) devices. Building upon this, recent studies have focused on perpendicularly magnetized magnetic tunnel junctions (p-MTJs), which offer improved scalability and thermal stability, making them more suitable for high-density memory applications. For example, a 2015 study by Huang *et al*. [238] showed that incorporating heavy metals like Ta or Pt significantly enhances spin-orbit coupling, reducing the critical current density for STT switching.

Voltage-controlled magnetic anisotropy (VCMA) in ultrathin CoFeB layers has emerged as a promising innovation for low-power memory applications. In 2022, Shao *et al* [240], demonstrated sub-volt switching in nanoscale voltage-controlled perpendicular magnetic tunnel junctions (p-MTJs), achieving a high VCMA coefficient (~130 fJ/Vm) and tunnel magnetoresistance (>150%). They reported sub-nanosecond precessional switching in 50–70 nm MTJs with voltages below 1V and scalability down to 30 nm MTJs at ~2V, highlighting the potential of VCMA-MRAM for low-power, high-density computing applications. These advancements underscore the transformative potential of material innovations in scaling MRAM technology.

**Racetrack Memory Technology: A Novel Spintronic Storage Paradigm** - Racetrack memory represents a revolutionary approach to spintronic storage, relying on the controlled manipulation of domain walls (DWs) within ferromagnetic nanowires [241]. A schematic diagram of the racetrack memory concept is shown in Figure 13 (a-d). Unlike conventional storage devices, which depend on stationary bits, racetrack memory stores data in the positions of DWs, moving them using spin-polarized currents. This solid-state approach eliminates mechanical components, ensuring durability and energy efficiency.

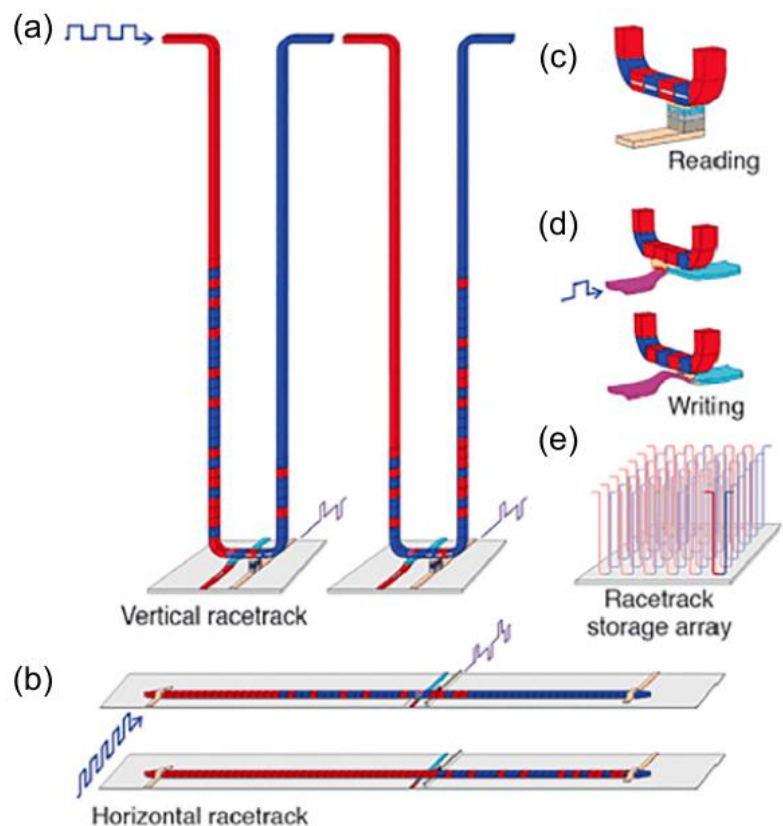

Fig 13. **Schematic of racetrack memory.** (a) Vertical and (b) horizontal configurations. (c) Electrical reading and (d) writing operations. (e) 3D-array architecture. Figure reprinted from [241].

Permalloy (NiFe) and Heusler alloys have emerged as key materials for racetrack memory due to their optimal DW mobility and low pinning [242]. In a pivotal 2011 study by Miron *et al* [243], demonstrated spin-transfer and spin-orbit torques, enhanced by the Rashba effect, enable fast and stable domain-wall motion in ultrathin Co nanowires with structural inversion asymmetry (SIA). The Rashba field stabilizes Bloch domain walls, overcoming Walker breakdown and depinning issues, allowing for higher mobility and domain-wall velocities up to 400 m/s, advancing memory and logic applications. These advancements highlight the importance of material design in ensuring reliable data transport.

Materials with strong DMI have further expanded the capabilities of racetrack memory. A 2016 study by Yang *et al*. [244], showcased the use of Pt/Co/Ir trilayers, where DMI-stabilized chiral DWs exhibited efficient current-driven motion. This development underscores the potential of interfacial engineering in optimizing spintronic devices.

Emerging materials such as ferrimagnetic GdFeCo have gained attention for their ability to achieve fast DW velocities with reduced Joule heating. In 2023, Ogawa *et al*. [245] used ultrafast MOKE imaging to study current-induced domain wall motion in Pt/GdFeCo films, revealing time-varying velocities influenced by Joule heating and spin torque effects, with DW acceleration observed ~1 ns after current injection. In 2018, Caretta *et al*. [132] demonstrated that current-induced domain wall motion in Pt/Gd$_{44}$Co$_{56}$/TaO$_x$ films can achieve velocities up to 1.3 km/s near the angular momentum compensation temperature (TA). marking a significant milestone in the field. Additionally, the integration of topological insulators like Bi$_2$Se$_3$ with ferromagnetic layers has shown promise in enhancing spin current efficiency, as evidenced by experimental results in 2022 that reported improved spin-transfer efficiency and reduced energy consumption [246]. They demonstrated that the spin Hall angle in Bi$_2$Se$_3$ can be effectively tuned asymmetrically and enhanced by about 600% through the application of a bipolar electric field, highlighting the potential of interfacial engineering in optimizing spintronic devices.

Both MTJs and racetrack memory illustrate the transformative potential of advanced magnetic materials in modern spintronic technologies. By combining cutting-edge materials science with innovative device architectures, these technologies are paving the way for scalable, high-performance, and energy-efficient memory solutions for next-generation computing.

## 1.6 Spintronic Nano-Oscillators: Fundamentals and Applications

Spintronic nano-oscillators (SNOs) are transformative devices in spintronics, exploiting the magnetization dynamics of nanoscale magnetic materials to generate high-frequency oscillations. First proposed in the early 2000s, these devices utilize STT or the SHE [247] to induce sustained magnetization precession in magnetic layers. Unlike conventional electronic oscillators, SNOs offer unparalleled energy efficiency, frequency tunability, and nanoscale integration, making them ideal for applications in communication technologies, neuromorphic computing, and signal processing. A significant milestone was the experimental demonstration of STT-driven oscillations by Kiselev *et al*. in 2003 [248], which validated the theoretical predictions of sustained magnetization dynamics under spin-polarized currents.

At the core of SNOs is the interaction between spin currents and magnetic moments. In STT-based SNOs, spin-polarized currents flow through MTJs or GMR stacks, transferring angular momentum to the magnetic layer and inducing precession. A key advancement in these devices came with the use of CoFeB/MgO multilayers [237], which offered high TMR ratios, enhancing signal output. CoFeB/MgO-based SNOs have demonstrated oscillation frequencies ranging from hundreds of megahertz to tens of gigahertz [249]. Additionally, research from Carpentieri *et al* [250] explored SNOs using perpendicular anisotropy in CoFeB/MgO-based

MTJs, further highlighting the potential of these structures for high-frequency applications. Similarly, SHE-based devices employ heavy metals such as Pt, Ta, or W to generate spin currents via spin-orbit coupling, eliminating the need for spin-polarized injection layers. Demidov *et al.* in 2012 [251] showcased SHE-driven oscillations in nano-gap geometry with NiFe/Pt bilayers, paving the way for simpler, more scalable devices.

One of the most promising features of SNOs is their ability to achieve synchronization. Mutual synchronization of multiple oscillators through dipolar coupling, spin-wave interactions, or electrical feedback enhances coherence and output power, which is crucial for practical applications [252, 253]. In 2017, Awad *et al.* [254] demonstrated synchronized spin Hall nano-oscillator (SHNO) arrays with reduced phase noise, achieving linewidths as low as 2 MHz. This breakthrough underscored the potential of SNOs for microwave communication systems, where signal stability is paramount. Another technique, injection locking, has been used to further stabilize oscillations, with studies reporting reduced linewidths in STT oscillators when locked to an external reference signal [255, 256].

Spintronic nano-oscillators are particularly suited for neuromorphic computing due to their ability to mimic neural network behavior. In 2018, Romera *et al.* [257] demonstrated that coupled SNO arrays could perform vowel recognition tasks, effectively emulating spiking neural networks. These devices operate at low power and are inherently scalable, making them attractive for hardware-embedded artificial intelligence. Moreover, SNOs have been utilized to realize hardware Ising machines for solving combinatorial optimization problems. For example, Albertsson *et al.* (2021) [258] used SNO arrays to solve max-cut graph problems, showcasing their potential in unconventional computing paradigms.

The choice of materials is critical in optimizing SNO performance. CoFeB/MgO remains a benchmark for STT oscillators, while NiFe/Pt [259] and CoFeB/Ta [260] bilayers are preferred for SHNOs due to their efficient spin-charge interconversion. Advanced architectures, including nano-constrictions, nanowires, and vortex oscillators [261], offer additional degrees of freedom in tuning oscillation frequencies and enhancing signal strength. Vortex-based oscillators, which utilize magnetic vortices as dynamic elements, have exhibited high-quality factors and are particularly suitable for generating low-phase-noise signals. In 2007, Pribiag *et al*. [262] demonstrated vortex oscillators with frequency modulation capabilities, expanding their potential for use in wireless communication systems. Emerging materials like TIs (e.g., $Bi_2Se_3$) and 2D magnetic materials (e.g., $CrI_3$) are also being explored to enhance SNO functionality. TIs provide robust spin-momentum locking, improving spin current generation and reducing energy losses. For instance, research on ferromagnetic insulators (FMIs) coupled with TI surface states has demonstrated nonlinear magnetization dynamics, where spin-polarized currents mediate phase-locked oscillations [263]. These coupled FMI-TI systems require low power and offer a pathway toward novel spintronic computing architectures. Additionally, hybrid systems combining SNOs with photonic or phononic elements [264] are being developed for quantum technologies, enabling applications such as microwave-to-optical transduction and magnon-photon coupling.

The future of SNOs lies in their ability to integrate seamlessly into existing electronic and photonic systems. Their unique combination of nanoscale dimensions, high-frequency operation, and low power consumption positions them as key components for next-generation technologies, from 5G communication to hardware-accelerated artificial intelligence. By advancing material science and device engineering, SNOs are poised to revolutionize both conventional and unconventional computing landscapes [265].

 **Next generation unconventional computing**

The progress of spintronic neuromorphic materials has been driven by advancements in material science and device functionality. Early studies introduced MTJs as artificial synapses, demonstrating magnetic memory-based synaptic behaviour. Subsequent breakthroughs in MgO-based MTJs and VCMA have improved synaptic precision, energy efficiency, and low-power operation, paving the way for next-generation neuromorphic computing architectures. Borders *et al.* fabricated MTJ-based synaptic arrays capable of online learning [266]. In their study, MTJ-based neural networks with 20,000 MTJs integrated on CMOS chips were demonstrated, closely resembling market-ready STT-MRAM technology. Using 36 dies, the study revealed that even minor defects in mapped networks significantly degrade performance unless accounted for during training. These devices showed improved stability and scalability, setting the stage for large-scale neuromorphic chips. The exploration of skyrmions for neuromorphic devices further advanced the field. In 2020, Pinna *et al.* [267] demonstrated the potential of computing using random skyrmion textures, highlighting the complex dynamical responses of these topological structures for pattern recognition tasks. Additionally, recent studies have explored the use of two-dimensional (2D) magnetic materials, such as $CrI_3$-based van der Waals heterostructures, to develop reconfigurable spintronic devices for compact neuromorphic applications [268].

The concept of reservoir computing (RC) has also gained traction in spintronics. In 2019, Riou *et al.* [269] showcased the use of spintronic nano-oscillators (SNOs) for RC, where their inherent nonlinear dynamics and synchronization capabilities enabled pattern recognition tasks. Similarly, Sun *et al.* [270] extended the idea to skyrmion-based reservoirs, leveraging their topological stability and chiral dynamics for energy-efficient data processing. The application of Brownian reservoir motion [271] in spintronic systems is emerging as another novel approach, where stochastic magnetic fluctuations in materials like ferrimagnets provide a basis for unconventional computation.

In 2024 YIG-based hybrid magnonic-photonic systems [272] further expanded the potential of spintronic neuromorphics. These systems enable microwave-to-optical signal conversion, bridging the gap between classical and quantum neuromorphic architectures. The introduction of retinomorphic computing [273], inspired by the human retina, employs 2D magnetic materials to mimic synaptic and neuronal behaviors, offering a compact and reconfigurable platform for image recognition and signal processing.

These milestones underscore the transformative role of innovative materials and device architectures in shaping the future of neuromorphic computing. From MTJs and skyrmions to hybrid magnonic-photonic systems and Brownian reservoirs, advancements in spintronic materials are paving the way for energy-efficient, scalable, and next-generation computational paradigms.

## 1.7 Future Research Directions

**Challenges and Opportunities -** The field of spintronic materials continues to evolve, but several challenges remain in unlocking their full potential. One critical issue is the optimization of material properties to minimize energy dissipation while maintaining high efficiency. For example, reducing damping in materials and developing room-temperature antiferromagnets with efficient spin transport are significant hurdles. Additionally, achieving scalable synthesis of complex materials, such as van der Waals magnets and topological insulators, is essential for integrating these materials into practical devices. The stability of spin textures, such as skyrmions and domain walls in real-world conditions including thermal noise and fabrication defects presents another challenge.

Enhancing interfacial effects like the DMI and exploring hybrid material systems could offer potential solutions. Moreover, materials like Pt/Co/Ta multilayers in spin-orbit torque (SOT) devices require further refinement to reduce critical switching currents and improve overall device reliability.

From a technological standpoint, ensuring compatibility with existing semiconductor technologies and developing scalable fabrication methods are crucial for widespread adoption. Spintronic devices must overcome limitations in readout speed and signal-to-noise ratios to compete with traditional electronic components. Integrating spintronic components into hybrid systems—such as spin-photon and spin-phonon devices—also presents challenges related to material compatibility and device architecture.

Despite these challenges, numerous opportunities exist. The discovery of new material systems, including 2D magnets like $CrI_3$ and Weyl semimetals, provides an exciting platform for exploring novel spintronic phenomena. Advances in nanofabrication techniques and computational modelling are enabling the design of devices with unprecedented precision, paving the way for entirely new spintronic applications.

**Potential Breakthroughs -**

The future of spintronic technology holds the potential to revolutionize various fields, ranging from memory and logic to unconventional computing and quantum technologies. One anticipated breakthrough is the development of all-spin logic circuits, where information is processed entirely through spin currents, eliminating the need for charge transport. This innovation could lead to ultra-low-power computing systems with drastically reduced energy consumption.

Another promising area is the exploration of non-Hermitian spin systems, where gain and loss mechanisms create exceptional points, enabling unidirectional spin-wave propagation. These systems could revolutionize spin-based signal processing and communication by offering new ways to control and manipulate spin waves.

The rise of quantum spintronics marks a convergence between spintronics and quantum computing. Materials such as topological insulators and superconductors offer pathways for quantum information storage and manipulation. Hybrid devices that integrate spintronic and photonic components may facilitate the development of quantum interconnects, bridging the gap between classical and quantum systems.

Additionally, advances in neuromorphic computing using spintronic materials are expected to provide energy-efficient solutions for artificial intelligence (AI) applications. Magnetic skyrmions, spintronic nano-oscillators, and hybrid magnonic systems are set to enable new paradigms in data processing and pattern recognition.

**1.8 Conclusion**

**Summary of Key Points and Outlook on Spintronic Device Technology -** Spintronic materials and devices have emerged as transformative elements in modern electronics, bridging fundamental physics with technological innovation. MTJs and SOT materials have laid the foundation for energy-efficient memory and logic devices, while novel structures such as racetrack memory and skyrmion-based technologies demonstrate the versatility of magnetic materials in unconventional computing.

Advancements in materials—such as 2D magnets, ferrimagnets, and hybrid magnonic-photonic systems—highlight the synergy between materials science and spintronic functionality. These innovations have enabled breakthroughs in neuromorphic computing, quantum interconnects, and low-power data storage solutions, underscoring the potential of spintronics to shape the future of technology.

Looking ahead, the development of new material systems, combined with advances in nanofabrication and computational techniques, will be crucial in overcoming current challenges. The integration of spintronic devices with existing semiconductor technologies, alongside the exploration of hybrid systems, promises to broaden the impact of spintronic devices across diverse domains.

As research continues to push the boundaries of what is possible, spintronic devices are poised to play a central role in meeting the growing demand for energy-efficient, scalable, and multifunctional technologies. Their significance in advancing computing, communication, and quantum systems underscores their transformative potential in reshaping the future of electronics.